\documentclass[12pt]{article}
\usepackage{amsmath,amsfonts,amssymb}
\usepackage{color}
\usepackage{epic,eepic}
\usepackage{theorem,cite}
\numberwithin{equation}{section}

\definecolor{brique}{rgb}{.9,.2,0}
\definecolor{blvert}{rgb}{0,.8,.85}
\definecolor{vertcl}{rgb}{0,1,.7}
\newcommand\vertcl[1]{\textcolor{vertcl}{#1}}
\newcommand\blvert[1]{\textcolor{blvert}{#1}}
\newcommand\brique[1]{\textcolor{brique}{#1}}
\def\lapth{
\begin{picture}(164,70)(0,-15)\thicklines
\put(0,0){\vertcl{\rule{20pt}{4pt}}}
\put(19,1){\vertcl{\line(1,3){23}}} 
\put(20,1){\vertcl{\line(1,3){23}}} 
\put(21,1){\vertcl{\line(1,3){23}}}
\put(22,1){\vertcl{\line(1,3){23}}}
\put(45,70){\vertcl{\line(1,-3){23}}} 
\put(44,70){\vertcl{\line(1,-3){23}}} 
\put(43,70){\vertcl{\line(1,-3){23}}}
\put(42,70){\vertcl{\line(1,-3){23}}}
\put(2,24){\vertcl{\rule{120pt}{4pt}}}
\put(65,0){\vertcl{\rule{60pt}{4pt}}}
\put(5,37){\Huge{\brique{\textbf{L}}}} 
\put(62,37){\Huge{\brique{\textbf{PTh}}}}
\put(12,-8){\blvert{\rule{92pt}{3.5pt}}}
\put(24,-15){\blvert{\rule{57pt}{3.5pt}}}
\put(36,-22){\blvert{\rule{30pt}{3.5pt}}}
\end{picture}
\raisebox{35pt}{
\begin{minipage}{320pt}\begin{center}
\textbf{Laboratoire d'Annecy-leVieux de Physique
Th\'eorique}\\[4ex]
website: \texttt{http://lappweb.in2p3.fr/lapth-2005/}
\end{center}
\end{minipage}}\\
\vspace{10pt}\quad \hrulefill\\
\vspace{10pt}}

\newcommand{\be}{\begin{equation}}
\newcommand{\ee}{\end{equation}}
\newcommand{\bea}{\begin{eqnarray}}
\newcommand{\eea}{\end{eqnarray}}
\newcommand{\beano}{\begin{eqnarray*}}
\newcommand{\eeano}{\end{eqnarray*}}
\newcommand{\nonu}{\nonumber \\}

\newcommand{\hs}[1]{\hspace{#1 mm}}

\def\cA{{\cal A}}                 
                 
       \def\cH{{\cal H}}          
                 \def\cL{{\cal L}}
       \def\cN{{\cal N}}

\newcommand{\atopn}[2]{\genfrac{}{}{0pt}{}{#1}{#2}}
\def\qmbox#1{\qquad\mbox{#1}\quad}

\newcommand{\wh}[1]{\widehat{#1}}

\newcommand{\mb}[1]{\hs{4}\mbox{#1}\hs{4}}
\newcommand{\half}{\frac{1}{2}}

\newcommand{\CC}{\mbox{${\mathbb C}$}}
\newcommand{\RR}{\mbox{${\mathbb R}$}}

\newcommand{\ZZ}{\mbox{${\mathbb Z}$}}

\newcommand{\II}{\mbox{${\mathbb I}$}}

\newcommand{\EE}{\mbox{${\mathbb E}$}}

\newcommand{\MM}{\mbox{${\mathbb M}$}}

\begin{document}
\renewcommand{\thefootnote}{\fnsymbol{footnote}}
\newpage
\pagestyle{empty}
\setcounter{page}{0}
\hspace{-1cm}\lapth

\vspace{20mm}

\begin{center}

{\LARGE  {\sffamily Exactly solvable 
 models in atomic and molecular physics}}\\[1cm]

{\large A. Foerster$^1$ and E. Ragoucy$^2$}\footnote{Emails: angela@if.ufrgs.br,
ragoucy@lapp.in2p3.fr}\\[.42cm] 
$^{1}$ Instituto de F\'{\i}sica da UFRGS \\ 
Av. Bento Gon\c{c}alves 9500, Porto Alegre, RS - Brazil
\vspace{0.5cm}\\
$^{2}$ Laboratoire de Physique Th{\'e}orique LAPTH\\[.242cm]
   LAPP, BP 110, F-74941  Annecy-le-Vieux Cedex, France. 
\end{center}
\vfill\vfill

\begin{abstract}
    We construct integrable generalised models in a systematic way exploring 
    different representations of the $gl(N)$ algebra. The models are then interpreted in 
    the context of atomic and molecular physics, most of them related to different 
    types of 
    Bose-Einstein condensates.
    The spectrum of the models is given through the analytical Bethe ansatz method.
    \\
    We further extend these results to the case of the superalgebra 
    $gl(M|N)$, providing in this way models which also include fermions.
\end{abstract}

\vfill
\vfill
\rightline{\tt cond-mat/0702129}
\rightline{LAPTH-1175/07}
\rightline{February 2007}

\newpage
\pagestyle{plain}
\setcounter{footnote}{0}
\section{Introduction}

Exactly solvable models are a fascinating issue that continue to attract 
considerable interest in physics and mathematics.
Altough the integrability of quantum systems is usually restricted to 
one dimension, there are many reasons that turn this study relevant
for physical applications. 
It serves as a test for computer analysis and analytical methods for realistic systems to 
which, until now, only numerical calculations and perturbative methods may be applied.
In addition, a nontrivial solvable model reveals the essence of the 
phenomena under consideration. For instance, many concepts established in critical phenomena were inspired
by the exact solution of the Ising model. 
{From} the experimental point of view, there 
are some real spin-1 compounds (e.g. NENC, NDPK, or NBYC etc\ldots)\cite{comp1}
and strong coupling ladder compounds (such as 
($5$IAP)$_2$CuBr$_4$$\cdot$$2$H$_2$O, or 
Cu$_{2}$(C$_5$H$_{12}$N$_2$)$_2$Cl$_4$, etc\ldots)\cite{prl} that can 
be perfectly well described by integrable models. The necessity of using
exactly solvable models has been also demonstrated through experimental 
research on aluminium grains at nanoscale level \cite{rbt}.

A significant aspect of integrable systems is its interdisciplinary character,
i.e., they can be found in different areas of physics.
The Ising and the Heisenberg models \cite{baxter1} in statistical mechanics, the 
t-J and Hubbard models \cite{fabkorep} in condensed matter physics, 
the nonlinear $\sigma$-model \cite{bytsko} in quantum field theory, 
the interacting Boson model \cite{iachello} in nuclear physics and more recently the 
two-site Bose-Hubbard model \cite{legget} in atomic and molecular 
physics are just some representative examples of the 
high impact and potentiality of these systems. Let us remark also the 
emergence of integrable systems in high energy physics, more 
particularly in gauge theories \cite{lip,FaKo,Zarembo} (for a 
recent review, see \cite{BBGK}), or string theory, through the recent analysis 
in super-Yang-Mill theories, see e.g. \cite{miza,beisstau}.

Therefore, new exactly solvable models are highly welcome and constitute the 
main focus of the present article. In particular we will concentrate 
on the construction of integrable generalised models 
in atomic and molecular physics, most of them related to 
Bose-Einstein condensates (BECs).

\null

The phenomenon of Bose-Einstein condensation, while predicted long ago \cite{ago}, 
is currently one of the most active fields in physics, responsible for many of  
new perspectives on the potential applications of quantum systems.
Since the early experimental realizations of BECs using ultracold 
dilute alkali gases \cite{early}, intense efforts have been devoted to the study of 
new properties of BEC. In recent years the creation of a molecular 
BEC form an atomic BEC has been obtained by different techniques \cite{mol}.
The field was further broadened by the achievement of quantum degeneracy 
in ultracold fermionic gases \cite{reffermi}. These achievements could lead to 
 new scientific investigations that includes 
coherent atomic lasers, quantum chemistry, 
the quantum gas with anisotropic dipolar interactions,  
quantum information, atomtronics, among many others.

In this context, it is natural to expect that exactly solvable models in the BEC scenario
may be of relevance,
providing some physical insights \cite{hertier}.
Our main purpose here is to employ the integrable systems machinery in its full power, 
i.e., exploring all possible types of representations of some algebra
(we consider, in particular the $gl(N)$ algebra) to enlarge the family of known 
exactly solvable models in atomic and molecular physics
with the aim that potentially new relevant models emerge.
Using this machinery
some existing models in the BEC scenario,
such as the two-site Bose-Hubbard model \cite{legget}
will be restored as 
well as new ones will be obtained.
Remarkably, a two-coupled BEC model with a field, a two-coupled BEC-model with different
types of atoms and a three-coupled BEC model, among others, 
will be introduced in this general framework. 
It is worth to mention here that the popular "BEC-transistor" in atomotronics 
uses a BEC in a triple well \cite{transistor}.
The models are then solved by means of Bethe ansatz methods.

Our paper is organized as follows: in section 2 we briefly review the general setting 
of integrable systems and fix notation.
In section 3 we present our general approach and also discuss the different 
representations of the $gl(N)$ algebra that will be adopted. Section 4 is devoted to the 
discussion of the different physical models we can get using this construction.
In section 5 the Bethe ansatz equations of the models are derived.
We extend these results to the case of the superalgebra 
$gl(M|N)$ in section 6 and some applications of this formalism, i.e. models 
which also include fermions are presented in section 7.
Section 8 is devoted to some concluding remarks.

\section{Generalities}
\subsection{Monodromy and transfer matrices}
We remind here the general setting used in the context of integrable spin 
chains and more generally in QISM \cite{QISM,QISM2, Kul, KulSkly}. One starts with a so-called 'algebraic' monodromy matrix $T(u)$, 
which is a $N\times N$ matrix taking values in an algebra $\cA$:
$$
T(u)=\sum_{i,j=1}^N T_{ij}(u)\,E_{ij}
\mb{with} T_{ij}(u)\in\cA\ ;\ E_{ij}\in End(\CC^N)
$$
where $E_{ij}$ are the $N\times N$ elementary 
matrices (with 1 at position $(i,j)$ and 0 elsewhere).
In the main part 
of the present paper, the algebra $\cA$ will be the Yangian of $gl(N)$, $Y(N)$,
but we 
will also study super-Yangians. 

The monodromy matrix
obeys the so-called FRT relation \cite{FRT}:
\begin{equation}
R_{12}(u-v)\,T_{1}(u)\,T_{2}(v) = T_{2}(v)\,T_{1}(u)\,R_{12}(u-v)
\label{rtt}
\end{equation} 
where we have used the standard auxiliary space notation, e.g. 
$T_1(u)=T(u)\otimes \II_N$, where $\II_N$ is the identity matrix 
and $R_{12}(u)$ is the $R$ matrix of $\cA$. The $R$-matrix obeys the Yang-Baxer equation
\begin{equation}
R_{12}(u_1-u_2)\,R_{13}(u_1-u_3)\,R_{23}(u_2-u_3)\,=
R_{23}(u_2-u_3)\,R_{13}(u_1-u_3)\,R_{12}(u_1-u_2)\,.
\end{equation}
It will also be  unitary
\begin{equation}
R_{12}(u_1-u_2)\,R_{21}(u_2-u_1)= f(u_1,u_2)\,\II\otimes\II
\label{eq:unit}
\end{equation}
where $f(u_1,u_2)$ is some known function.

It is  well-known that one can produce 
monodromy matrix for several sites by applying the coproduct $\Delta T(u) =
T(u)\otimes T(u) \equiv T^{[1]}(u)\,T^{[2]}(u)$ where the superscript 
labels in which copy of the algebra $T(u)$ acts. More generally, one 
can consider 
$$
T(u)=T^{[1]}(u)\,T^{[2]}(u)\ldots\,T^{[L]}(u)
$$
as a monodromy matrix. This is the base of
spin chain models, the copies of the algebra defining (upon
representation) the $L$ quantum spaces (sites) of the chain, see e.g. 
\cite{korepin, faddeev,KulResh,ow} and references therein. In what follows,
we will call $T^{[n]}(u)$ the elementary monodromy matrices, the product of
all these elementary matrices providing the 'real' monodromy matrix.

Let us stress that to fix a physical model, one has to represent the 
monodromy matrix, i.e. assign to each of the $T^{[n]}(u)$ a representation 
of the algebra $\cA$: changing the representations will lead to different 
physical models.

Once we have a (represented) monodromy matrix $T(u)$, then (\ref{rtt})
ensures that the transfer 
matrix $t(u)=tr T(u)$ obeys
\begin{equation}
{[t(u),t(v)]}=0\,.
\label{eq:integrab}
\end{equation}
 Expanding $t(u)$ in the variable $u$ leads to 
commuting integral of motions, one of them being the chosen Hamiltonian of 
the system.  For
instance, in spin chain models, one takes this Hamiltonian to be
$\cH=t(0)^{-1}\,t'(0)$, while e.g. in the Bose-Hubbard model, it is simply $t(0)$. 
The other quantities constructed from the transfer matrix just produce
conserved quantities. If the number of such (independent) conserved quantities
is sufficiently large, the system is said to be integrable.

\subsection{Automorphisms of the monodromy matrices}
We present some automorphisms of the relation (\ref{rtt}) that will be 
of some use in the physical models we will study.

The first automorphism is built on the transposition: starting from a 
monodromy matrix $T(u)$, it is easy to show that 
$$
T(u)\ \to\ T^t(-u) \mb{i.e.} T_{ij}(u)\ \to\ T_{ji}(-u)
$$
where the transposition is done in the auxiliary space. The proof 
relies on the unitary relation (\ref{eq:unit}) of the $R$-matrix.
We will call this automorphism the sign-transposition.

Another automorphism is the conjugation by a constant matrix
$$
T(u)\ \to\ M\,T(u)\,M^{-1} \mb{with}M\in End(\CC^{N})
$$
which is a consequence of the invariance of the $R$-matrix
$$
M_{1}\,M_{2}\,R_{12}(x) = R_{12}(x)\,M_{1}\,M_{2}
$$
We will call this automorphism a conjugation.

A particular case of conjugation is the dilatation automorphism
$$
T(u)\ \to\ \alpha\,T(u) \mb{with} \alpha\in\CC
$$

\subsection{Hermiticity}
 The elementary monodromy matrices we will consider below
are always hermitian
\begin{equation}
\Big(T(u)\Big)^\dag=T(u) \mb{i.e.} T^{\dag}_{ij}(u)=T_{ji}(u)
\end{equation}
This implies that
\begin{equation}
\Big(T(u)\Big)^\dag=\Big(T^{[1]}(u)\,T^{[2]}(u)\Big)^\dag
=T^{[2]}(u)\,T^{[1]}(u)
\end{equation}
so that the total monodromy matrix is not hermitian. However, using cyclicity 
of the trace, we get:
\begin{equation}
t^\dag(u)=
tr\Big(T^{[2]}(u)\,T^{[1]}(u)\Big)=tr\Big(T^{[1]}(u)\,T^{[2]}(u)\Big)=t(u)
\end{equation}
Thus, the transfer matrix is hermitian. This property is valid only when $L=2$ 
(and $L=1$), cyclicity being not sufficient to get hermiticity of the transfer 
matrix as soon as $L\geq3$. For this reason, we will focus below on the case 
$L=2$, hence ensuring hermitian Hamiltonians. We shall see that even with
this restriction, we will get most of the models used in the BEC context, as 
well as new ones.

\section{Bosonic $gl(N)$ models}
We present here the general approach we use, focusing on the case of the 
 Yangian $Y(N)=Y(gl(N))$ \cite{Drinfeld}. Other cases are presented in the sections below.
 The $R$-matrix we consider takes 
the form \cite{baxter1,baxter2,yang}:
\begin{equation}
R_{12}(x) =
\II\otimes\II-\frac{1}{x}\,P_{12}, 
\label{RmatY}
\end{equation}
where 
$$
P_{12}=\sum_{i,j=1}^N E_{ij}\otimes E_{ji}
$$
 is the permutation operator.

\subsection{Conserved quantities}
Since $L=2$, the monodromy and transfer matrices have an expansion 
\begin{eqnarray}
T_{kl}(u) &=& \sum_{n=0}^2 u^n\,T_{kl}^{(n)}
\mb{with} T_{kl}^{(2)}=\omega_k\,\delta_{kl}\mb{and}\omega_k\in\CC
\label{eq:expansT}\\
t(u) &=& t_2\, u^2 + t_1\, u + t_0 \mb{with} t_2\in\CC
\end{eqnarray}
For BEC models, one generally uses $t_0=t(0)$ as an Hamiltonian, 
while $t_1$ and $t_2$ correspond to integrals of motion.
It is easy to see that $t_2$ is just a number, but from the
explicit form of the $R$-matrix, one can get other conserved quantities. 

 Indeed starting from the relation (\ref{rtt}) and projecting on the basis 
elements $E_{ij}\otimes E_{kl}$ in the auxiliary spaces, one gets:
\begin{equation}
 {[T_{ij}(u)\,,\,T_{kl}(v)]} = \frac{1}{u-v}\ \Big(
 T_{kj}(u)T_{il}(v)-T_{kj}(v)T_{il}(u)\Big)\,.
\end{equation}
Then, taking $i=j$, summing on $j$, and looking at the coefficient of $v$,
one gets
\begin{equation}
 {[t(u)\,,\,T_{kl}^{(1)}]} = (\omega_k-\omega_l)\,
\big(T_{kl}^{(0)}+u\,T_{kl}^{(1)}\big)
\end{equation}
This proves in particular that the quantities
\begin{equation}
I_k=T_{kk}^{(1)}\ ,\ \forall\ k=1,...,N
\label{eq:consvI}
\end{equation}
commute with the transfer matrix and are in involution. 
Thus, they generate integrals of motions.

Let us remark that, following the value of the $\omega_k$ parameters, 
one could get more conserved quantities (through the $T_{kl}^{(1)}$, 
$k\neq l$, generators), but they will not form an abelian subalgebra: 
they will generate a symmetry algebra for the model.

\null

\subsection{Spin chain monodromy matrices \label{sect:spinCh}}
The elementary monodromy matrices for the Yangian 
can be specialized to $gl(N)$ 
monodromy matrices using the so-called evaluation map. This amounts to
take these elementary monodromy matrices to be of the form 
\begin{equation}
    L(u) = \sum_{i,j=1}^{N} L_{ij}(u)\,E_{ij} \mb{with}
    L_{ij}(u)=u\delta_{ij}+ e_{ij}
\label{eq:matL}
\end{equation}
or in matricial form
\begin{equation}
L(u)=\left(\begin{array}{ccccc}
u+ e_{11} &  e_{12} &  e_{13} & \ldots &  e_{1N} 
\\
e_{21} & u+ e_{22} &  e_{23} & \ldots &  e_{2N} 
\\
\vdots & \ddots & \ddots  & \ddots & \vdots \\
 e_{N-2,1} & &    &  &  e_{N-1,N}  \\
 e_{N1} &  e_{N2} & \ldots &  e_{N,N-1} & u+ e_{NN}
\end{array}\right)
\end{equation}
Here, $e_{ij}$ are $gl(N)$ \textit{unrepresented} generators obeying
\begin{equation}
 {[}e_{ij}\,,\,e_{kl}] = \delta_{jk}\,e_{il}- \delta_{il}\,e_{kj}
\end{equation}
It is easy to show that $L(u)$ obey the relation (\ref{rtt}) with 
the $R$-matrix (\ref{RmatY}). Hermiticity of $L(u)$ is ensured
by $e_{ij}^\dag=e_{ji}$. Moreover, one can use the Yangian shift automorphism
$u\to u+w$ to get extra free parameters. Thus,
$$T(u)=L^{[1]}(u+w_1)\,L^{[2]}(u+w_2)$$ leads to hermitian integrable models
with transfer matrix
\begin{equation}
    t(u)=\sum_{j=1}^{N}(u+w_{1}+e_{jj}^{[1]})\,\otimes\,(u+w_{2}+e_{jj}^{[2]}) 
    \ +\ \sum_{j\neq k}^{N}e_{jk}^{[1]}\,\otimes\,e_{kj}^{[2]}\,. 
\end{equation}
In the context of spin chain models, the parameters $w_{j}$ are 
called inhomogeneity parameters.

As already stated, it is the choice of a $gl(N)$ representation
 for each of the sites that will determine the physical model one wishes to
work on. When the representations are highest weight finite dimensional ones, 
it leads to spin chains models.
They have been extensively studied and 
we just repeat here well-known facts to illustrate the techniques we shall 
use with different representations.

For instance, one can take the fundamental representation of $gl(N)$
$$
\pi(e_{ij})=E_{ij}\ ,\ i,j=1,...,N
$$ 
for both elementary monodromy matrices, leading to a (well-known and somehow trivial) 
two-site spin chain. Specifying furthermore to the case of $gl(2)$, 
one recovers the Pauli matrices 
$$
\pi(e_{12})=\sigma_+\quad ;\quad \pi(e_{21})=\sigma_-\quad ;\quad 
\pi(e_{11}-e_{22})=\sigma_z\quad ;\quad \pi(e_{11}+e_{22})=\II_2
$$
leading to an Hamiltonian
$$
H=t(0)=\sigma_+\otimes\sigma_- + \sigma_-\otimes\sigma_+
 + \half\,\sigma_z\otimes\sigma_z +\half
$$
Of course, one could choose another representation, 
for instance, for $gl(2)$, take the spin 1 representation
\begin{eqnarray}
&& \pi(e_{12})=\left(\begin{array}{ccc} 
0 & 1 & 0\\ 0 & 0 & 1 \\ 0 & 0 & 0
\end{array}\right)=S_+\quad ;\quad 
\pi(e_{21})=\left(\begin{array}{ccc} 
0 & 0 & 0\\ 1 & 0 & 0 \\ 0 & 1 & 0
\end{array}\right)=S_-\quad ;\quad
S_{z}=\left(\begin{array}{ccc} 
1 & 0 & 0\\ 0 & 0 & 0 \\ 0 & 0 & -1
\end{array}\right)\qquad\qquad\\
&&\pi(e_{11})=\left(\begin{array}{ccc} 
1 & 0 & 0\\ 0 & \half & 0 \\ 0 & 0 & 0
\end{array}\right)=\half(S_z+\II_3)\quad ;\quad
\pi(e_{22})=\left(\begin{array}{ccc} 
0 & 0 & 0\\ 0 & \half & 0 \\ 0 & 0 & 1
\end{array}\right)=\half(\II_3-S_z)
\end{eqnarray}
leading to the Hamiltonian
$$
H=t(0)=S_+\otimes S_- + S_-\otimes S_+
 + \half\,S_z\otimes S_z +\half
$$
Note that in both cases, the parameters $w_n$ do not play any role because 
the spin chain is too simple. For the same reason, the above 
Hamiltonians coincide with $t(0)^{-1}\,t'(0)$.

\subsection{Oscillator monodromy matrices}
The above framework can be generalized to other (infinite dimensional) 
representations of $gl(N)$. 

\subsubsection{Bosonic and fermionic representations of $gl(N)$}
We start with $N$ couples of creation/annihilation operators
$(a_{i},a^\dag_{i})$, with commutation relations
\begin{equation}
 {[a_{i},a^\dag_{j}]}=\mu_{i}\,\delta_{ij}\quad;\quad  
 {[a^\dag_{i},a^\dag_{j}]}={[a_{i},a_{j}]}=0\,.
\end{equation}
{From} these relations, it is straightforward to check that
$\cL(u)$ defined by
\begin{equation}
 \cL(u)=\sum_{i,j=1}^{N} \cL_{ij}(u)\,E_{ij} \mb{with} 
 \cL_{ij}(u) =
 \mu_{i}\,u\,\delta_{ij}+\frac{q_{i}}{q_{j}}\,a^\dag_{i}\,a_{j}
 \end{equation}
 obey the relations
\begin{equation}
 {[\cL_{ij}(u)\,,\,\cL_{kl}(v)]} = \frac{1}{u-v}\ \Big(
 \cL_{kj}(u)\cL_{il}(v)-\cL_{kj}(v)\cL_{il}(u)\Big)\,.
 \end{equation}
 It is equivalent to
\begin{equation}
R_{12}(u-v)\,\cL_{1}(u)\,\cL_{2}(v) = 
\cL_{2}(v)\,\cL_{1}(u)\,R_{12}(u-v)
\qmbox{with} R_{12}(x) = \II\otimes\II-\frac{1}{x}\,P_{12}
\end{equation} 
so that $$T(u)=\cL^{[1]}(u+w_{1})\cL^{[2]}(u+w_{2})$$
provides an integrable model.

In fact, this calculation is valid for an arbitrary number 
of sites\footnote{Of course, we will potentially lose hermiticity 
of the transfer matrix when $L\geq3$.}, 
and it 
just corresponds to a choice of (infinite dimensional) $gl(N)$
representation 
$$
\pi(e_{ij})=a_i^\dag\,a_j\ ,\ i,j=1,...,N
$$
for the elementary monodromy matrices (\ref{eq:matL}). 
The highest weight is the Fock space vacuum $|0>$, but the representation is 
reducible and is an infinite sum of finite dimensional representations with 
fixed 'particle number' $\cN=\sum_i a^\dag_i\,a_i$. 
We will call the corresponding monodromy matrix an 'homogeneous oscillator 
monodromy matrix'.

Focusing on hermitian elementary matrices, one is led to 
$$
\mu_{j}\in\RR \mb{and} \vert q_{i}\vert^2 = \vert q_{j}\vert^2
\quad\forall i,j
$$
The last equation imposes 
$$
q_{j} = q_{0}\,e^{i\theta_{j}}\mb{with} q_{0},\theta_{j}\in\RR
$$
The parameter $q_{0}$ is irrelevant for $\cL(u)$, and since 
$$
a_{j} \to e^{i\theta_{j}}\,a_{j} \mb{and}
a^\dag_{j} \to e^{-i\theta_{j}}\,a^\dag_{j}
$$
is an invariance of the algebra, one can restrict to the case
\begin{equation}
 \cL_{ij}(u) =
 \mu_{i}\,u\,\delta_{ij}+ a^\dag_{i}\,a_{j}\mb{with}\mu_{j}\in\RR
\end{equation}
or in matricial form
\begin{equation}
\cL(u)=\left(\begin{array}{ccccc}
\mu_{1}\,u+ n_{1} &  a_{1}^\dag\,a_{2} &  a_{1}^\dag\,a_{3} 
& \ldots & a_{1}^\dag\,a_{N} 
\\
 a_{2}^\dag\,a_{1} & \mu_{2}\,u+ n_{2} & a_{2}^\dag\,a_{3} 
& \ldots & a_{2}^\dag\,a_{N} 
\\
\vdots & \ddots & \ddots  & \ddots & \vdots  \\
 a_{N-2}^\dag\,a_{1} & &    &  & a_{N-1}^\dag\,a_{N}  \\
 a_{N}^\dag\,a_{1} & a_{N}^\dag\,a_{2} & \ldots 
& a_{N}^\dag\,a_{N-1} & \mu_{N}\,u+ n_{N}
\end{array}\right)
\mb{with} n_{i}=a_{i}^\dag\,a_{i}
\end{equation}
In general, one takes the values $\mu_j=1$ to get canonical commutation relations. 
Then, $\cL(u)$ has a leading term (in $u$) which is just the identity matrix, 
in accordance with the definition of the Yangian matrix $L(u)$.

Let us remark that there exists also a fermionic $gl(N)$ representation 
where now 
$a_j$ and $a_j^\dag$ are fermionic operators:
\begin{equation}
 \cL^f_{ij}(u) =
 \mu_{i}\,u\,\delta_{ij}+ a^\dag_{i}\,a_{j}\mb{with}\mu_{j}\in\RR
\end{equation}
In that case, the representation is finite dimensional, 
 since the oscillators now obey the supplementary 
relations $(a_j)^2=0=(a_j^\dag)^2$. This possibility will be used 
to produce some
fermionic models.

\subsubsection{Inhomogeneous oscillator monodromy matrices}

The above calculation is valid whatever
the values of the numbers $\mu_{i}$ are.
In particular, one can take the value $\mu_{j}=0$, for some
$j\in J\subset [1,N]$: the corresponding $\cL(u)$ matrix will still obey 
(\ref{rtt}) with the $R$-matrix (\ref{RmatY}). This particular value
$\mu_j=0$ allows a (scalar) representation
$a_{j}=\alpha_{j}\in\CC$ and $a^\dag_{j}=\alpha_{j}^{*}$,  $j\in J$, of
the oscillator algebra. For obvious reason, we will call these operators 
'constant oscillators', and 
'inhomogeneous oscillator monodromy matrix' the corresponding elementary 
monodromy matrix. To distinguish it from the homogeneous one, when needed, 
we will denote it as $\Lambda(u)$ instead of $\cL(u)$.

Hence, $\Lambda(u)$ is an $N\times N$ matrix built on $p=N-|J|$ couples
$(a_{j},a_{j}^\dag)$, $p$ being independent from $N$ (provided it is smaller
than $N$). Then, $T(u)$ will lead to a transfer matrix based $p_1+p_2$
oscillators, where $p_n$ is the number of oscillators in 
$T^{[n]}(u)\equiv \Lambda^{[n]}(u)$, 
$n=1,2$. 

Let us remark that since we have taken the limit $\mu_j\to0$ for $j\in J$, 
the elementary monodromy matrices do not start with $\II_N$, but rather 
with a non-invertible diagonal matrix. In that sense, we are not in the 
Yangian context anymore. However, since the relation (\ref{rtt}) is still 
obeyed with the $R$-matrix (\ref{RmatY}), this does not affect the relation
(\ref{eq:integrab}), so we are still in the framework of 
hermitian integrable models. The underlying algebraic structure, which is 
very close to the Yangian, was studied in \cite{molinge} and 
is called 'truncated Yangians'.
Keeping in mind this restriction, 
we will lose keep writing that we are in the 
Yangian context.

Note also that the limit $\mu_j\to0$ (and $a_j,a_j^\dag$ constant) 
can be taken at the very end of the calculations. Hence, we can consider
 a general $\cL(u)$ matrix,
keeping in mind that, to get a $\Lambda(u)$ elementary monodromy matrix, and 
depending on the model one wishes to study, some of the $a_j$, 
$a_j^\dag$ operators will be in fact complex numbers $\alpha_j$, $\alpha^*_j$, 
and the 
corresponding $\mu_j=0$. 
For other oscillators, one chooses in general $\mu=\pm1$.

Finally, let us remark that for the fermionic $gl(N)$ representation,
since now the
$a_j$ and $a_j^\dag$ operators obey the 
supplementary 
relations $(a_j)^2=0=(a_j^\dag)^2$, it is not possible to take them 
as non-vanishing 
constants. Taking all the constants to be zero leads to trivial models,
hence, fermions are excluded from $\Lambda(u)$ when dealing with $Y(N)$. 
Fortunately, we will see 
below that one can recover them when studying models based on super-Yangians 
(see section \ref{sect:super}).

\subsubsection{Automorphisms of oscillator algebra}
%

An automorphism will be used to produce new terms in the Hamiltonians. 
It exists only for the bosonic algebra, and consists in a shift by a constant:
\begin{eqnarray}
(a,a^\dag,\mu) &\to& (a+\alpha,a^\dag+\alpha^*,\mu)\ ,\ \alpha\in\CC
\mb{for bosons}
\label{eq:shift}
\end{eqnarray}
We will call this automorphism a shift of the oscillator algebra. 
It can be used to produce boundary terms in the different models.

\subsection{$gl(N)$ transfer matrices}
We have seen that we have essentially two types of elementary matrices 
at our disposal, 
the matrices
$L(u)$ and $\cL(u)$, so that one gets three types of transfer matrix:
\begin{eqnarray}
t(u) &=& tr\left( L^{[1]}(u+w_1)\,L^{[2]}(u+w_2) \right)\nonu
&\sim&  u\,\sum_{i=1}^N\big(E^{[1]}_{ii}
+ E^{[2]}_{ii}\big) 
+\sum_{i,j=1}^N E^{[1]}_{ij}\, E^{[2]}_{ji}
+\sum_{i=1}^N\big(w_2\,E^{[1]}_{ii}
+w_1\, E^{[2]}_{ii}\big) \\
t(u) &=& tr\left( L^{[1]}(u+w_1)\,\cL^{[2]}(u+w_2) \right)\nonu
&\sim&  u\,\sum_{i=1}^N\big(\mu_i\,E^{[1]}_{ii}
+ n_{i}\big) 
+\sum_{i,j=1}^N E^{[1]}_{ij} a^\dag_{j}a_{i}
+\sum_{i=1}^N\big(w_2\,E^{[1]}_{ii}
+w_1\, n_{i}\big) \\
t(u) &=& tr\left( \cL^{[1]}(u+w_1)\,\cL^{[2]}(u+w_2) \right)\nonu
&\sim& u\,\sum_{i=1}^N\big(\mu_i\,n_{bi}
+ \nu_i\,n_{ai}\big) 
+\sum_{i,j=1}^N b^\dag_{i} a^\dag_{j}a_{i}b_{j}
+\sum_{i=1}^N\big(w_1\,n_{bi}
+ w_2\,n_{ai}\big)
\end{eqnarray}
where the $\sim$ sign means equality modulo polynomials in $u$ with 
constant coefficients.
As a notation, we have introduced $E^{[n]}_{ij}=\pi_n(e_{ij})$, $n=1,2$, 
the representation of the $gl(N)$ generators in $L^{[n]}(u)$ and 
called $a_j,a^\dag_j,\mu_j$ 
(resp. $b_j,b^\dag_j,\nu_j$)
the oscillator algebras in $\cL^{[1]}(u)$ (resp. in $\cL^{[2]}(u)$); $n_{aj}$ 
(resp. $n_{bj}$) are the corresponding number operators.

The first transfer matrix is just a two-site spin chain, where the first sites 
carries 
the 'spin' $\pi_1$ of $gl(N)$, and the second site the 'spin' $\pi_2$. 
These models (and
their generalization to an arbitrary number of sites)
have been studied for a long time, and we will not consider them here. 
The two other transfer matrices lead to several physical models, 
depending on the choices of:
\begin{itemize}
\item The $gl(N)$ algebra one considers (i.e. the choice of $N$)
\item The $gl(N)$ representation in the $L(u)$ part
\item The characteristic (bosonic or fermionic) of the oscillators in the $\cL(u)$ part
\item The number of 'constant oscillators' in the $\Lambda(u)$ part(s)
\item The values (specially zero or not) of the parameters corresponding 
to these 'constant oscillators'.
\item The use or not of the different automorphisms.
\end{itemize}
The next section is devoted to the presentation of the different physical 
models one can get from these choices.

\section{Examples of $gl(N)$ BEC models\label{sect:bos}}
For simplicity, we now focus on the case where the parameters 
involved in the elementary matrices are real, and normalize the 
commutator of the oscillators to 1. More general 
Hamiltonians (still hermitian) can be obtained keeping complex 
parameters, as detailed above.
\subsection{Models based on two by two matrices}
The possible elementary monodromy matrices take the form
\begin{eqnarray}
L(u) &=& \left(\begin{array}{cc} u+\half\,S_z & S_+ \\ 
S_- & u-\half\,S_z \end{array}\right)
\qquad
\cL(u) =
\left(\begin{array}{cc} u+n_1 & a_1^{\dag}\,a_2 \\ 
 a_2^{\dag}\,a_1 & u+n_2 \end{array}\right)
\label{eq:L2x2-1}\\
\Lambda(u) &=& 
\left(\begin{array}{cc} u+n & \beta\,a^{\dag} \\ 
\beta\,a & \beta^2 \end{array}\right)
\qquad \wh\Lambda(u) = 
\left(\begin{array}{cc} -\beta^2 & \beta\,a^{\dag} \\ 
\beta\,a &  u-n \end{array}\right) 
\label{eq:L2x2-2}
\end{eqnarray}
plus possibly the use of shift automorphisms. We have used the 
sign-transposition and dilation automorphisms
$$
\wh\Lambda(u)\ \to\ -\wh\Lambda^t(-u)
$$
plus a redefinition of the $\beta$ parameter to make 
$\wh\Lambda(u)$ similar to $\Lambda(u)$.

Apart from the spin chain model presented in section \ref{sect:spinCh}, 
one gets 5 different models. 
The conserved quantities $I_{j}$ have the general form:
\begin{eqnarray}
I_1 = \mu_{b1}\,n_{a1}+\mu_{a1}\,n_{b1}\mb{and}
I_2 = \mu_{b2}\,n_{a2}+\mu_{a2}\,n_{b2}
\end{eqnarray}
where $(a_{1},a_{2})$ refer to the first elementary monodromy matrix, 
and $(b_{1},b_{2})$ to the second one.

\subsubsection{A spin-boson model}
We consider $t(u)=tr L(u+w_1)\,\Lambda(u+w_2)$. 
Up to irrelevant constant terms, 
it takes the form
$$
t(u) \equiv u(\half\,S_z+n) 
+ w_1(\half\,S_z+n)
+ ( \alpha\,S_+a+\alpha\,S_-a^\dag )
+ \half S_z\,n 
\label{spinboson}
$$
leading to Hamiltonian $H=t(0)$ with conserved quantity
\begin{eqnarray}
I  = n + \half S_z 
\end{eqnarray}
Above $a^\dag$ $(a)$ denotes the single-mode field creation (annihilation) operator, 
$S_z, S_{\pm}$ the atomic inversion, rising and lowering operators and $w_1$ is the transition frequency.
This model describes the interaction of a two-level atom with a single-mode radiation field. 
It was derived in \cite{kundu5}, \cite{dukk}  using different methods and it reduces to the 
Jaynes-Cummings Hamiltonian on resonance and in the rotating-wave approximation 
in the absence of the last term \cite{jc}. Despite of its simplicity, this model 
has been a source of insight into a better comprehension of the nuances of the interaction 
between light and matter. It is important to remark that a Jaynes-Cummings model interaction 
can be experimentally realized in cavity-QED setups and also, as an effective interaction 
in laser cooled trapped ions \cite{unicamp}.

Again, the shift automorphism produces boundary term
\begin{eqnarray}
 H_{bound}&=& \beta\Big(
\half S_z(a^\dag+a)
+\alpha\,(S_++S_-)
+w_1(a^\dag+a)\Big)
\end{eqnarray}
to the transfer matrix. The conserved quantity is modified to
\begin{eqnarray}
I'  &=& n + \half S_z 
+\beta\,(a^\dag+a)
\end{eqnarray}

\subsubsection{Generalised spin-boson model}
We consider $t(u)=tr L(u+w_1)\,\cL(u+w_2)$. Up to irrelevant constant terms, 
it takes the form
$$
t(u) \equiv u(n_1 +n_2) 
+ \half S_z(n_1 -n_2)
+S_+a_1^\dag a_2+S_-a_2^\dag a_1 
+w_1(n_1 +n_2)
$$
leading to Hamiltonian 
\begin{equation}
H=t(0)=\half S_z(n_1 -n_2)
+S_+a_1^\dag a_2+S_-a_2^\dag a_1 
+w_1(n_1 +n_2)
\end{equation}
with conserved quantities
\begin{eqnarray}
I_1  = n_1 + \half S_z \ ;\ 
I_2 = n_2- \half S_z 
\end{eqnarray}
Above the oscillators $a_{j}$, $j=1,2$ denote two radiation 
fields (two photons, for example)
interacting with a two-level atom. 
The atom-field interacting term could be interpreted as a scattering of two 
fields with a two-level atom. Here we mention that 
if linearly polarised light is used, it is possible
to have the same transition frequency $w_1$ (see, for example \cite{rego}).

When the oscillators are bosonic, one can use the shift automorphism to add
a boundary term:
\begin{eqnarray}
H_{bound}&=& \alpha_1\Big(
\half S_z(a^\dag_1+a_1+\alpha_1)
+S_+a_2+S_-a_2^\dag
+w_1(a^\dag_1+a_1)\Big)\nonu
&+& \alpha_2\Big(
\half S_z(a^\dag_2+a_2+\alpha_2)+
S_+a_1+S_-a_1^\dag
+w_1(a^\dag_2+a_2)\Big)+ \alpha_1\,\alpha_2\,(S_++S_-)
\end{eqnarray}
to the transfer matrix. The conserved quantities then become
\begin{eqnarray}
I'_1  &=& n_1 + \half S_z 
+\alpha_1\,(a^\dag_1+a_1)\\ 
I'_2 &=& n_2- \half S_z
+\alpha_2\,(a^\dag_2+a_2)
\end{eqnarray}

\subsubsection{Simple heteroatomic-molecular BEC model\label{sec:simplheteroBEC}}
We consider $t(u)=tr \cL^{[1]}(u+w_1)\,\cL^{[2]}(u+w_2)$, with 
$a_j,a^\dag_j$, $j=1,2$ for $\cL^{[1]}$ and 
$b_j,b^\dag_j$, $j=1,2$ for $\cL^{[2]}$.

Up to irrelevant constant terms, 
the transfer matrix takes the form
$$
t(u) \sim u(n_1 +n_2) 
+ n_{a1} \,n_{b1}+ n_{a2}\,n_{b2}
+a_1^\dag b_2^\dag b_1 a_2
+a_2^\dag b_1^\dag b_2 a_1 
+w_1\,n_b +w_2\,n_a
$$
where we have introduced the notation
\begin{eqnarray}
&&n_{a1}=a_1^\dag a_1 \ ;\ 
n_{b1}=b_1^\dag b_1 \ ;\ 
n_{a2}=a_2^\dag a_2\ ;\ 
n_{b2}=b_2^\dag b_2\\
&& n_1 = n_{a1}+ n_{b1}\ ;\
n_{2} = n_{a2}+ n_{b2} \\
&& n_a = n_{a1} + n_{a2}\ ;\ 
n_b = n_{b1} + n_{b2}\,.
\end{eqnarray}
It leads to Hamiltonian $H=t(0)$ with conserved quantities
\begin{eqnarray}
I_1  = n_1 \ ;\ 
I_2 = n_2
\end{eqnarray}
This Hamiltonian is a particular case of the one presented in the 
section \ref{sect:atomolBEC}, and we postpone the physical discussion 
to this section.

%
Finally, using the quantum determinant (see appendix), one can also show from 
$c_2$ given in appendix that $n_{a}n_{b}$ is also conserved, so that one finally gets as 
conserved quantities
$$
n_1 = n_{a1}+ n_{b1}\ ;\
n_{2} = n_{a2}+ n_{b2} \ ;\
n_a = n_{a1} + n_{a2}\ ;\ 
n_b = n_{b1} + n_{b2}
$$
three of them being independent.

\subsubsection{Heteroatomic-molecular BEC model\label{sect:atomolBEC}}
We consider $t(u)=tr \cL(u+w_1)\,\Lambda(u+w_2)$, with 
$a,a^\dag$ and $b,b^\dag$ for $\cL(u)$ and 
$c,c^\dag$ for $\Lambda(u)$.

Up to irrelevant constant terms, 
it takes the form
$$
t(u) \equiv u(n_a +n_c) 
+ n_a \,n_c+ \beta^2\,n_{b}
+\beta\,(b^\dag c^\dag a + a^\dag b c)
+w_1\,n_c +w_2\,n_a 
\label{heteroatomic}
$$
It leads to Hamiltonian $t(0)$ with conserved quantities
\begin{eqnarray}
I_1  = n_a+n_b \ ;\ 
I_2 = n_a+n_c
\end{eqnarray}
It is trivial to check that this Hamiltonian (adding terms $I_{1}^{2}$ 
and $I_{2}^{2}$) corresponds to the heteroatomic-molecular
Bose-Einstein condensate model\cite{jzrg}
\begin{eqnarray}
{\cal H} &=& U_{aa}\,n_a^2 + U_{bb}\,n_b^2 +U_{cc}\,n_c^2
+U_{ab}\,n_a\,n_b+U_{ac}\,n_a\,n_c+U_{bc}\,n_b\,n_c 
\nonumber \\
 &+& \mu_a\,n_a + \mu_b\,n_b+\mu_c\,n_c 
 +\Omega\,(a^{\dag}bc +c^{\dag}b^{\dag}a)\,,
\label{ham-heterromol}
\end{eqnarray}
for the particular choice of the couplings 
$\half U_{aa}=U_{bb}=U_{cc}=U_{0}+U_{1}$, $U_{ab}=4\,U_{0}$, 
$U_{bc}=0$, $U_{ac}=4\,U_{1}+1$, 
$\mu_a=w_2$, $\mu_b=\beta^2, \mu_c =w_1$, $\Omega = \beta$.

In this context, the parameters $U_{ij}$ describe S-wave scattering, 
$\mu_i$ are external potentials and $\Omega$ is the amplitude for interconversion 
of atoms and molecules.
One gets a three-mode Hamiltonian describing a 
Bose-Einstein condensate with two distinct species of atoms, denoted $b$ and $c$, 
which can combine to produce a molecule $a$ \cite{jzrg}. The total atom number $(I_1+I_2)$ and the 
imbalance between the atomic modes $(I_1-I_2)$ are conserved quantities.
A detailed classical and quantum analysis of this model reveals unexpected 
scenarios, such as the emergence of quantum phases 
when the imbalance is zero \cite{maj}.

%
\subsubsection{The two-site Bose-Hubbard model}
We consider $t(u)=tr \Lambda^{[1]}(u+w_1)\,\Lambda^{[2]}(u+w_2)$, with 
$a,a^\dag$ for $\Lambda^{[1]}$ and $b,b^\dag$ for $\Lambda^{[2]}$.

Up to irrelevant constant terms, 
it takes the form
$$
t(u) \equiv u(n_a +n_b) 
+ n_a \,n_b+ \omega\,(b^\dag a + a^\dag b )
+w_1\,n_b +w_2\,n_a
\label{bosehub}
$$
with
$ w = \alpha \beta $. 
It leads to Hamiltonian $t(0)$ with conserved quantity
\begin{eqnarray}
I  = n_a+n_b \,.
\end{eqnarray}
It is easy to verify that by combining the conserved quantity $I$ with
the Hamiltonian $t(0)$ and choosing properly the coupling constants, 
we arrive at\footnote{More
specifically, we have $ H = c ( I^2 - 4t(0) )$, where the following 
identification has been done: $K/8=c$ ; $(\Delta \mu)/2=4 c w_2= - 4 c w_1$
; ${\cal {E} _J}/2 = 4 c w$.}
\bea
{\cal H}&=& \frac {K}{8}\,(n_a- n_b)^2 - \frac{\Delta \mu}{2}\, (n_a -n_b)
 -\frac {\cal{E}_J}{2}\, (a^\dagger b + b^\dagger a).
\label{ham-2Hub} 
\eea
This is the two-site Bose-Hubbard model, also known as the canonical 
Josephson Hamiltonian\cite{legget}. 
It describes the tunneling between two single particle states or 
modes ($a$ and $b$), which can 
be separated spatially (two wells) or internally (two different
internal quantum numbers). The parameter $K$ corresponds to the
atom-atom interaction, 
$\Delta \mu$ is the external potential
and  $\cal {E} _J$ is the coupling for the tunneling. 
Despite of its apparent simplicity, this model predicts 
the existence of a threshold coupling between a delocalised 
and self-trapped phase\cite{milb},\cite{our}, in qualitative agreement with 
experiments\cite{albiez}.

\subsection{Models based on three by three matrices\label{sect:3x3}}
The number of possibilities increases very fast, we present here 
only some cases that we found physically relevant. The interested 
reader can easily compute the other models using the techniques we have 
described.
\subsubsection{Two-coupled BEC model with a single-mode field\label{sect:BECfield}}
We consider the elementary monodromy matrices
\begin{eqnarray}
 \Lambda^{[1]}(u) &=& \left(\begin{array}{ccc}
 u+n_{a} & a^\dag\,b & \alpha_{3}\,a^\dag\\ 
 b^\dag\,a & u+n_{b} & \alpha_{3}\,b^\dag\\ 
 \alpha_{3}\,a & \alpha_{3}\,b & \alpha_{3}^2
\end{array}\right)\mb{and}
\Lambda^{[2]}(u)=\left(\begin{array}{ccc}
 u+n_{c} & \beta_{1}\,c^\dag & \beta_{2}\,c^\dag\\ 
 \beta_{1}\,c & \beta_{1}^2 & \beta_{1}\,\beta_{2}\\ 
 \beta_{2}\,c & \beta_{1}\,\beta_{2} & \beta_{2}^2
\end{array}\right)\quad\nonumber\\[1.2ex]
T(u) &=& \Lambda^{[1]}(u+w_1)\,\Lambda^{[2]}(u+w_2)
\end{eqnarray}
Then, the transfer
matrix reads
\begin{eqnarray}
t(u) &\sim& (n_{c}+n_{a})\,u
+\beta_{1}^2\,n_{b}+n_{a}\,n_{c}
+ w_{1}\,n_{c}+w_{2}\,n_{a}
+\alpha_{3}\,\beta_{2}\,(a^\dag\,c+c^\dag\,a)
\nonu
&& +\beta_{1}\,(a^\dag\,b\,c+b^\dag\,c^\dag\,a)
+\alpha_{3}\,\beta_{1}\,\beta_{2}\,(b^\dag+b)
\end{eqnarray}
with conserved quantity
$$ I=n_{a}+n_{c}.$$
This leads to the Hamiltonian\footnote{The following identification has been 
done:
$\beta_1 = w$ ; $\mu_a=w_2$ ; $\mu_c=w_1$ ; $\Omega = \alpha_3 \beta_2$.}
\begin{eqnarray}
{\cal H} = \mu_a\,n_{a} + \mu_c\,n_{c} + w^2 \, n_{b} 
+ \Omega\,( a^\dag\,c + c^\dag\,a) 
+ w\,( a^\dag\,b\,c +  c^\dag\,b^\dag\,a )
+ \Omega\,w\,(b^\dag + b )
\end{eqnarray}
which  describes two wells ($A$ and $C$) with atoms interacting 
with a single-mode field. 
Above $b^\dag$ and $b$ denote the single field creation  and annihilation operators, respectively 
and $a^\dag$ and $a $ ($c^\dag$ and $c $)
the creation and annihilation operators for a particle in the well 
$A$ $(C)$.
The total number of particles $n = n_a + n_c$, where $n_a = a^\dag a $ and $n_c = c^\dag c $ is conserved.
The parameter $\Omega$ is the coupling for the tunneling between the two wells, $w$ is the radiation field frequency and
$\mu_a, \, \mu_c$ are the external potentials.
Here there are basically two mechanisms that allow the atoms to trap between 
the wells: i) the tunneling effect, which is related with the depth of the wells;
ii) a tunneling which occurs mediated by a single-mode field $b$. 

When $w=0$, one has a supplementary
relation $$[t(u),n_{b}]=0, $$ and this Hamiltonian reduces to the two-site Bose-Hubbard (\ref{bosehub}).
When $\Omega=0$ we get 
$$[t(u),n_{b}-n_{c}]=0 $$ and this Hamiltonian reduces to (\ref{heteroatomic})

\subsubsection{Three coupled BEC model\label{mixed3}}
We consider the elementary monodromy matrices
\begin{eqnarray}
 \Lambda^{[1]}(u) &=& \left(\begin{array}{ccc}
u+n_{1} & a_{1}^\dag\,a_{2}& \beta_{3}\,a_{1}^\dag \\
a_{2}^\dag\,a_{1} & u+n_{2} & \beta_{3}\,a^\dag_{2} \\
\beta_{3}\,a_{1} & \beta_{3}\,a_{2} & \beta_{3}^2 \\
 \end{array}\right)\mb{and}
\Lambda^{[2]}(u)=\left(\begin{array}{ccc}
 -\beta_{1}^2 & \beta_{1}\,\beta_{2} & \beta_{1}\,a_3^\dag\\ 
 \beta_{1}\,\beta_{2} & -\beta_{2}^2 & \beta_{2}\,a_3^\dag\\ 
 \beta_{1}\,a_3 & \beta_{2}\,a_3 & u-n_3
\end{array}\right)\quad\nonumber\\[1.2ex]
T(u) &=& \Lambda^{[1]}(u+w_1)\,\Lambda^{[2]}(u+w_2)
\end{eqnarray}
where in $\Lambda^{[2]}(u)$ we have used the sign-transposition and 
dilatation automorphisms, and a redefinition of the parameters 
$\beta_{1}$ and $\beta_{2}$.
Then, the transfer
matrix reads
\begin{eqnarray*}
t(u) &\sim& \beta_1\,\beta_2\,(a_1^\dag a_2 + a_2^\dag a_1) 
+\beta_{3}\,\beta_1\,(a_1^\dag a_3 +a_3^\dag a_1)
+\beta_{3}\,\beta_2\,(a_2^\dag a_3 + a_3^\dag a_2)-
\beta_1^2\,n_1- \beta_2^2\,n_2- \beta_{3}^2\,n_3 \label{threewell}
\end{eqnarray*}
It leads to the Hamiltonian $t(0)$ 
with conserved quantity $ I = n_1+n_2+n_3$.
This corresponds to a special three coupled BEC model with asymmetric tunneling
and external potentials. To our knowledge, this is the first 
{\it integrable quantum model} describing a three well system\footnote{Notice, however, 
that it corresponds
to a simplified three-well model, with no atom-atom interaction terms.}.
We remark here that there has been recently an increasing interest in 
the study of three-well systems (trimers) for a variety of reasons, such as
\begin{itemize}
\item[\textit{i)}] its possible application in the
construction of a BEC-transistor \cite{transistor};
\item[\textit{ii)}] it is the simplest model which provides a bridge between the double-well and 
the multi-well systems \cite{milb5}, \cite{chin};
\item[\textit{iii)}] recent achievements in the experimental field, in particular 
the control promised by microtraps \cite{4} suggest the realization of the trimer to be 
at hand \cite{prlpena}.
\end{itemize}

It is easy to check that by combining the conserved quantity $I$ with
the Hamiltonian $t(0)$ and choosing properly the coupling constants, we arrive at
the following Hamiltonian\footnote{More specifically, $ {\cal H} = t(0) + \alpha n $ and we have 
done the change of variables
$\beta_1 \rightarrow \beta$ ; $\beta_3 \rightarrow \beta$ ; 
$\beta_2 \rightarrow \gamma/\beta$,
together with the 
following identification $ \Omega_2 = \gamma$ ; $\Omega = \beta^2$ ;
$\mu = (\alpha - \beta^2)$ ; $\mu_2 = (\alpha - \gamma^2/\beta^2)$.}
\begin{eqnarray}
{\cal H}&=& \Omega_2\,(a_2^\dag a_1 + a_1^\dag a_2 
+ a_2^\dag a_3 + a_3^\dag a_2 )
+  \Omega\, (a_1^\dag a_3 + a_3^\dag a_1) +
\mu \, n_1+ \mu \,n_3 + \mu_2 \,n_2 , \label{trimer0}
\end{eqnarray}
which describes an array of three coupled wells, 
which will be 
referred to as the left $(1)$, middle $(2)$ and right $(3)$ wells, respectively.
Above, $\Omega$ (resp. $\Omega_2$) denote the tunneling of atoms between 
the left and the right wells
(resp. the left-middle tunnelling and the middle-right tunneling), while 
$\mu_2$ and $\mu$ are the external potentials. Obviously, adapting the 
choice of parameters, one can also 
treat the case where the left-middle and middle-right tunnellings are 
different.

For the particular case where $\Omega$ approaches to zero this model reduces to the asymmetric 
open trimer model in the absence of the interatomic scattering and a large external potential $\mu_2$ \cite{prlpena}.
We observe here that if we consider this model in its full generality \cite{prlpena}, with 
$\Omega_2$ and $\mu_2$ as adjustable parameters, a BEC-transistor \cite{transistor}
can be derived.


In the symmetric limit $\Omega= \Omega_2 \,\, \,  \mu = \mu_2$ in
(\ref{trimer0}) we recover the model of the 
three coupled BEC based on the $SU(3)$ symmetry, proposed by Milburn et al. \cite{milb5} also in the 
absence of the interatomic scattering.

\subsubsection{Two-coupled BEC model with different types of atoms\label{sect:2cplBEC}}
We consider the elementary monodromy matrices
\begin{eqnarray}
 \Lambda^{[1]}(u) &=& \left(\begin{array}{ccc}
u+n_{a1} & a_{1}^\dag\,a_{2}& \alpha\,a_{1}^\dag \\
a_{2}^\dag\,a_{1} & u+n_{a2} & \alpha\,a^\dag_{2} \\
\alpha\,a_{1} & \alpha\,a_{2} & \alpha^2 \\
 \end{array}\right)\mb{and}
\Lambda^{[2]}(u)=\left(\begin{array}{ccc}
u+n_{b1} & b_{1}^\dag\,b_{2}& \beta\,b_{1}^\dag \\
b_{2}^\dag\,b_{1} & u+n_{b2} & \beta\,b^\dag_{2} \\
\beta\,b_{1} & \beta\,b_{2} & \beta^2 \\
\end{array}\right)\quad\nonumber\\[1.2ex]
T(u) &=& \Lambda^{[1]}(u+w_1)\,\Lambda^{[2]}(u+w_2)
\end{eqnarray}
Then, the transfer matrix reads  
\begin{eqnarray}
t(u) &\sim& u(n_a+n_b) + n_{a1}\,n_{b1} + n_{a2}\,n_{b2} 
+w_1\,n_b+w_2\,n_a + a_1^\dag b_2^\dag a_2 b_1+ a_2^\dag b_1^\dag a_1 b_2 \nonu
&& + \Omega \,(a_1^\dag b_1 +a_2^\dag b_2 +
b_1^\dag a_1 + b_2^\dag a_2) \label{difat}
\end{eqnarray}
where $\Omega = \alpha\beta$.
It leads to the Hamiltonian $t(0)$ 
with conserved quantities $n_j=n_{aj}+n_{bj}$, $j=1,2$.
This corresponds to a model of two wells ($A$ and $B$) with $n_j$ atoms 
of type $j$, $j=1,2$.
Here $n_{aj}$ ($n_{bj}$) denotes the number of atoms of type $j$, 
$j=1,2$ in the well $A$ ($B$).
Basically, this Hamiltonian describes the tunneling of 
atoms of different types ($1$ and $2$) in the two wells ($A$ and $B$).

Notice that this Hamiltonian could also be interpreted as describing two wells ($A$ and $B$)
with two levels ($1$ and $2$) in each well. Particles can tunnel between the wells and levels.
The tunneling term $\Omega$ allow particles to tunnel between wells just in the same level.
Similar models (four-mode Hamiltonians with tilted potentials) have been proposed 
recently in \cite{tilted}. In this context we mention that multi-mode models are 
receiving more attention, specially in connection to the creation of a quantum computer
from neutral atoms \cite{tilted}.

\subsubsection{Creation/dissociation of a molecule with two conformations\label{sect:conform}}
We consider the elementary monodromy matrices
\begin{eqnarray}
 \Lambda^{[1]}(u) &=& \left(\begin{array}{ccc}
u+n_{A1} & A_1^\dag\,a_{1}& \alpha\,A_1^\dag \\
a_{1}^\dag\,A_1 & u+n_{a1} & \alpha\,a^\dag_{1} \\
\alpha\,A_1 & \alpha\,a_{1} & \alpha^2 \\
 \end{array}\right)\mb{and}
\Lambda^{[2]}(u)=\left(\begin{array}{ccc}
-\beta^2 & \beta\,a_{2}^\dag &  \beta\,A_{2}^\dag \\
\beta\,a_{2} & u-n_{a2} & A_{2}^\dag\,a_{2}  \\
\beta\,A_{2} & a_{2}^\dag\,A_{2} & u-n_{A2}   \\
\end{array}\right)\quad\nonumber\\[1.2ex]
T(u) &=& \Lambda^{[1]}(u+w_1)\,\Lambda^{[2]}(u+w_2)
\end{eqnarray}
where we have used the sign-transposition and dilatation automorphisms 
on $\Lambda^{[2]}(u)$.
Then, the transfer
matrix reads
\begin{eqnarray}
t(u) &\sim& u(n_{a1}-n_{a2}) - n_{a1}\,n_{a2}  
-w_1\,n_{a2}+w_2\,n_{a1} -\alpha^2\, n_{A2}- \beta^{2}\,n_{A1}\nonu
&&+\alpha\beta\,(A_1^\dag A_2  + 
A_2^\dag A_1)  + \alpha\,(a_1^\dag a_2^\dag\, A_2 + A_2^\dag\, a_1 a_2)
+\beta\,(a_1^\dag a_2^\dag\, A_1 +A_1^\dag\, a_2 a_1) 
\end{eqnarray}
with conserved quantities $\delta n=n_{a2}-n_{a1}$ and
$n_{tot}=2\,(n_{A1}+n_{A2})+n_{a1}+n_{a2}$.

The Hamiltonian describes a molecule A, which exists in two 
conformations (two different stereochemical forms)
$A_{1}$ and $A_{2}$, and is constituted with two atoms (or 
submolecules) 
$a_{1}$ and $a_{2}$. There are transitions between the two
conformations 
of the molecule, and there is recombination/dissociation between the atoms and 
the two aspects of the molecule $A$. In this context, $\alpha$ and 
$\beta$ are related to the probabilities to obtain $A_{1}$ or $A_{2}$ 
starting from the atoms (or submolecules) $a_{1}$ and $a_{2}$.
The relative proportion of atoms $\delta n$ and the total number of 
atoms in the system $n_{tot}$ are 
conserved.

\subsubsection{A generalized heteroatomic-molecular BEC model\label{sect:glBEC}}
We consider the elementary monodromy matrices
\begin{eqnarray}
 \Lambda^{[1]}(u) &=& \left(\begin{array}{ccc}
u+n_{a1} & A^\dag\,a_{2}& \alpha\,A^\dag \\
a_{2}^\dag\,A & u+n_{a2} & \alpha\,a^\dag_{2} \\
\alpha\,A & \alpha\,a_{2} & \alpha^2 \\
 \end{array}\right)\mb{and}
\Lambda^{[2]}(u)=\left(\begin{array}{ccc}
u+n_{a1} & \beta\,a_{1}^\dag &  a_{1}^\dag\,b_{2} \\
\beta\,a_{1} & \beta^2 & \beta\,b_{2}  \\
b_{2}^\dag\,a_{1} & \beta\,b_{2}^\dag & u+n_{b2}   \\
\end{array}\right)\quad\nonumber\\[1.2ex]
T(u) &=& \Lambda^{[1]}(u+w_1)\,\Lambda^{[2]}(u+w_2)
\end{eqnarray}
Then, the transfer matrix reads
\begin{eqnarray}
t(u) &\sim& u(n_{A}+n_{a1}) + n_{A}\,n_{a1}  
+w_1\,n_{a1}+w_2\,n_{A} +\alpha^2\, n_{b2}+\beta^2\, n_{a2}
+ \beta\,(A^\dag a_1 a_2 +a_2^\dag a^\dag_1 A) \nonu
&&  + \alpha\beta\,(a_2^\dag b_2  + b_2^\dag a_2)
\end{eqnarray}
It leads to the Hamiltonian $t(0)$ with conserved quantity $n_1=n_{A}+n_{a1}$.
This model describes a molecule $A$ which can be dissociated into two different 
atoms $a_1$ and $a_2$. One of these atoms, $a_2$, can evolve to a 
different  state $b_2$, which forbids the recombination to $A$. For 
instance, $A$, $a_{1}$ and $a_{2}$ can be trapped in one well, the 
transitions between $a_{2}$ and $b_{2}$ corresponding to a `leak' of 
the $a_{2}$ atom toward a second well. 

In the limit $\beta \rightarrow 0 $ we recover the model (\ref{heteroatomic}).

\subsubsection{Coupling of two oscillators with $gl(3)$\label{sect:gl3}}
For completeness we give here an example of coupling $gl(3)$ with oscillators. We take the case
of two oscillators, but one can also couple a single oscillator, or three of them.
 \begin{eqnarray}
\Lambda(u) &=& \left(\begin{array}{ccc}
u+n_{1} & a_{1}^\dag\,a_{2}& \alpha\,a_{1}^\dag \\
a_{2}^\dag\,a_{1} & u+n_{2} & \alpha\,a^\dag_{2} \\
\alpha\,a_{1} & \alpha\,a_{2} & \alpha^2 \\
 \end{array}\right)
\mb{;}
L(u) = \left(\begin{array}{ccc}
u+e_{11} & e_{12} &  e_{13} \\
e_{21} & u+e_{22} & e_{23} \\
e_{31} & e_{32} & u+e_{33} \end{array}\right)\\[1.2ex]
T(u) &=& \Lambda(u)\,L(u)
\end{eqnarray}
The transfer matrix reads 
\begin{eqnarray}
t(u) &\sim& 
u\,(n_{1}+n_{2}+e_{11}+e_{22})+n_{1}\,e_{11}+n_{2}\,e_{22}
+a_{1}^\dag\,a_{2}\,e_{21}+a_{2}^\dag\,a_{1}\,e_{12}+\alpha^2\,e_{33}\nonu
&& +\alpha(a_{1}^\dag\,e_{31}+ a_{2}^\dag\,e_{32})+ 
\alpha(a_{1}\,e_{13}+ a_{2}\,e_{23})+w_1(e_{11}+e_{22})+ w_2(n_{1}+n_{2})
\end{eqnarray}
The conserved quantities are
$$ 
I_j=e_{jj}+n_j\ ,\ j=1,2
$$

\subsection{Coupling of one oscillator with $gl(N)$}
 We take 
\begin{eqnarray}
\Lambda(u) &=& \left(\begin{array}{cccc}
\mu\,u+n_{1} & \alpha_{2}\,a^\dag & \ldots & \alpha_{N}\,a^\dag \\
\alpha_{2}\,a & & & \\
\vdots & & \MM & \\
\alpha_{N}\,a & & & \end{array}\right)
\mb{;}
L(u)=\left(\begin{array}{cccc}
u+e_{11} & e_{12} & \ldots & e_{1N} \\
e_{21} & & & \\
\vdots & & u\,\II+\EE & \\
e_{N1} & & & \end{array}\right)\qquad\\[1.2ex]
\mbox{with} && \MM_{ij}=\alpha_{i}\,\alpha_{j} \mb{and}
\EE_{ij}=e_{ij}\nonumber\\[1.2ex]
T(u) &=& \Lambda(u)\,L(u)
\end{eqnarray}
Then, for $\mu=1$ and $\alpha$'s real, one gets
\begin{equation}
t(u)\sim u\,(e_{11}+n_{1}) +n_{1}\,e_{11}+\sum_{j=2}^{N}\left\{
\alpha_{j}\,\left(a^\dag\,e_{j1}+e_{1,j}\,a\right)+\alpha_{j}^2\,e_{jj}\right\}
+\sum_{2\leq j< k\leq N} \alpha_{j}\,\alpha_{k}(e_{jk}+e_{kj})
\end{equation}
We have not used the parameters $w_{j}$, but from the form of
$\Lambda(u)$ and $L(u)$, it is easy to see that one recovers them
through the shifts $n_{1}\to n_{1}+w_{1}$
and $e_{jj}\to e_{jj}+w_{2}$, $j=1,2,\ldots,N$.

\subsection{Two-coupled BEC model with $(N-1)$ levels}
We consider $T(u)=\Lambda^{[1]}(u+w_{1})\,\Lambda^{[2]}(u+w_{2})$ with
\begin{eqnarray}
 \Lambda^{[1]}_{jk}(u) &=& \delta_{jk}\,u+ a^\dag_{j}\,a_{k}\quad j,k<N
 \\
 \Lambda^{[1]}_{jN}(u) &=& \alpha_{N}\,a_{j}
 \mb{;} \Lambda^{[2]}_{Nj}(u) = \alpha_{N}\,a^\dag_{j}\quad j<N
 \mb{and} 
\Lambda^{[1]}_{NN}(u) = \alpha_{N}^2 \\
 \Lambda^{[2]}_{jk}(u) &=& \delta_{jk}\,u+ b^\dag_{j}\,b_{k}\quad j,k<N
 \\
 \Lambda^{[2]}_{jN}(u) &=& \beta_{N}\,b_{j}
 \mb{;} \Lambda^{[2]}_{Nj}(u) = \beta_{N}\,b^\dag_{j}\quad j<N
 \mb{and} 
\Lambda^{[2]}_{NN}(u) = \beta_{N}^2 
\end{eqnarray}
where $b_{j},b^\dag_{j}$ is another set of oscillators. The transfer
matrix $t(u)=trT(u)$ now reads
\begin{eqnarray}
t(u) &\sim& 
\sum_{j=1}^{N-1}\left\{u\,\left(n^{[2]}_{j}
+n^{[1]}_{j}\right)+n^{[2]}_{j}\,n^{[1]}_{j}+
\alpha_{N}\beta_{N}\,(a_{j}^\dag b_{j} + b_{j}^\dag a_{j})\right\} \nonu
&&+\sum_{1\leq j\neq k\leq N-1} a_{j}^\dag a_{k} b_{k}^\dag b_{j} 
\end{eqnarray}
where $n^{[1]}_{j}=a^\dag_{j}\,a_{j}$ and
$n^{[2]}_{j}=b^\dag_{j}\,b_{j}$. It leads to the `fundamental'
Hamiltonians
\begin{eqnarray}
t(0) &\sim& 
\sum_{j=1}^{N-1}\left(n^{[2]}_{j}
+n^{[1]}_{j}\right)\\
t'(0) &\sim& \sum_{j=1}^{N-1}\left\{ n^{[2]}_{j}\,n^{[1]}_{j}+
\alpha_{N}\beta_{N}\,(a_{j}^\dag b_{j} + b_{j}^\dag a_{j})\right\} 
+\sum_{1\leq j\neq k\leq N-1} a_{j}^\dag a_{k} b_{k}^\dag b_{j} 
\end{eqnarray}
Again, the shifts $n_{j}^{[p]}\to n_{j}^{[p]}+w_{p}$, $j=1,2,\ldots,N-1$,
$p=1,2$ give back the $w$ dependence.

It corresponds to a generalization of the previous Hamiltonian (\ref{difat}) , 
which could be interpreted, for example, as describing two wells with 
$N-1$ levels in each well. 
The quantities $I_{j}=n_{j}^{[1]}+n_{j}^{[2]}$, $j=1,2,\ldots, N-1$, 
are conserved.

\subsection{$N$-coupled BEC model}
We consider the elementary monodromy matrices
\begin{eqnarray}
 \Lambda^{[1]}_{jk}(u) &=& \delta_{jk}\,u+ a_{j}\,a^\dag_{k}\quad j,k=1,2
 \\
 \Lambda^{[1]}_{jk}(u) &=& \alpha_{k}\,a_{j} \mb{;}
 \Lambda^{[1]}_{kj}(u) = \alpha_{k}\,a_{j}^\dag\quad j=1,2\ ;\ k=3,...,N\\
\Lambda^{[1]}_{jk}(u) &=& \alpha_{k}\,\alpha_{j} \quad j,k=3,...,N\\
 \Lambda^{[2]}_{jk}(u) &=& \alpha_{k}\,\alpha_{j}\quad j,k=1,2\\
 \Lambda^{[2]}_{jk}(u) &=& \delta_{jk}\,u+ a^\dag_{j}\,a_{k}\quad j,k=3,...,N
 \\
 \Lambda^{[2]}_{jk}(u) &=& \alpha_{j}\,a_{k} \mb{;}
 \Lambda^{[2]}_{kj}(u) = \alpha_{j}\,a_{k}^\dag\quad j=1,2\ ;\ k=3,...,N
\end{eqnarray}
The transfer matrix takes the form
\begin{eqnarray}
t(u) &\sim& \half\,\sum_{i=1}^2 
\sum_{\atopn{j=1}{j\neq i}}^N \alpha_i\alpha_j\, (a_i^\dag a_j
+a_j^\dag a_i) +\sum_{i=1}^N \alpha_i^2\, n_i
\end{eqnarray}
It corresponds to a multi-well system, which generalizes the previous Hamiltonian (\ref{threewell}).
The total number operator $\sum_{i=1}^N n_i$ is conserved.

\section{Bethe ansatz equations}
We use the results of \cite{byebye}, applied to the special 
representations we are studying. This can be done when the 
representations are lowest weight. 
However, when the representations are reducible,
the Bethe ansatz will give the eigenvalues only on the irreducible parts. It has
to be applied for each lowest weights.

For the spin chain part, associated to the elementary monodromy matrix $L(u)$, 
one has always a unique lowest weight $v$ defining the representation 
(the spin) one is working with.
When dealing with oscillators, the Fock vacuum $|0>$
is a natural lowest weight, but different cases can appear, as we shall see.

\subsection{ $N=2$ case}
For $N=2$, the elementary monodromy matrices (\ref{eq:L2x2-1}) and (\ref{eq:L2x2-2}) 
become all triangular when applied to the Fock vacuum $|0>$ and/or to 
the lowest weight vector $v_{s}$:
\begin{eqnarray}
L(u)\,v_s &=& \left(\begin{array}{cc} u+s & S_+ \\ 
0 & u-s \end{array}\right)\,v_s\,,\ s\in\half\ZZ_{+}
\qquad
\cL(u)\,|0> =
\left(\begin{array}{cc} u & 0 \\ 
 0 & u \end{array}\right)\,|0>
\\
\Lambda(u)\,|0> &=& 
\left(\begin{array}{cc} u & \beta\,a^{\dag} \\ 
0 & \beta^2 \end{array}\right)\,|0>
\qquad \wh\Lambda(u)\,|0> = 
\left(\begin{array}{cc} \beta^2 & \beta\,a^{\dag} \\ 
0 &  u \end{array}\right)\,|0>
\end{eqnarray}
Note that $\cL(u)$ is proportional to the identity matrix, 
because the representation is reducible. Indeed all the vectors 
$|p>=(a^\dag_2)^p\,|0>$, $p\in\ZZ_+$, are lowest weight vectors for $\cL(u)$:
\begin{equation}
\cL(u)|p> =
\left(\begin{array}{cc} u+p & a_1^\dag\,a_2 \\ 
 0 & u \end{array}\right)\,|p>
\end{equation}
Then, depending on the model we are studying, we will get as pseudo-vacuum 
for the monodromy matrix, either the total Fock vacuum $|0,0>=|0>\otimes|0>$,
or different combinations of the type $|p,q>=|p>\otimes|q>$, 
$|v_s,p>=v_s\otimes|p>$, etc... When several pseudo-vacua are at our
disposal, we will have to repeat the Bethe ansatz method (described below)
for each of the pseudo-vacuum in order to get a complete set of
eigenvalues for the transfer matrix.

In all cases, the total monodromy matrix is triangular, 
and the pseudo-vacuum obeys:
\begin{eqnarray*}
T^{LL}_{jj}(u)\,|v_s,v_r> &=& \lambda_{j}(u) \,|v_s,v_r> 
\ ,\quad j=1,2 \mb{with} \left\{
\begin{array}{l} \lambda_{1}(u) = (u+w_1+s)(u+w_2+r) \\[1.2ex]
\lambda_{2}(u) =  (u+w_1-s)(u+w_2-r)\end{array}\right. 
\\[1.2ex]
T^{L\cL}_{jj}(u)\,|v_s,p> &=& \lambda_{j}(u)\,|v_s,p> 
\ ,\quad \left\{\begin{array}{l} j=1,2 \\ p\in\ZZ_{+}\end{array}\right. 
\mb{with} \left\{
\begin{array}{l} \lambda_{1}(u) = (u+w_1+s)(u+w_2+p) \\[1.2ex]
    \lambda_{2}(u) =  (u+w_1-s)(u+w_2)\end{array}\right. 
\\[1.2ex]
T^{\cL\cL}_{jj}(u)\,|p,q> &=& \lambda_{j}(u)\,|p,q> 
\ ,\quad \left\{\begin{array}{l} j=1,2  \\ p,q\in\ZZ_{+}\end{array}\right. 
\mb{with} \left\{
\begin{array}{l} \lambda_{1}(u) = (u+w_1+p)(u+w_2+q)\\[1.2ex]
   \lambda_{2}(u) = (u+w_1)(u+w_2) \end{array}\right. 
\end{eqnarray*}
\begin{eqnarray*}
T^{L\Lambda}_{jj}(u)\,|v_s,0> &=& \lambda_{j}(u)\,|v_s,0>
\ ,\quad j=1,2 \mb{with} \left\{
\begin{array}{l} \lambda_{1}(u) = (u+w_1+s)(u+w_2) \\[1.2ex]
    \lambda_{2}(u) = \alpha^2(u+w_1-s) \end{array}\right. 
\\[1.2ex]
T^{\cL\Lambda}_{jj}(u)\,|p,0> &=& \lambda_{j}(u)\,|p,0>
\ ,\quad \left\{\begin{array}{l} j=1,2 \\ p\in\ZZ_{+}\end{array}\right. 
\mb{with} \left\{
\begin{array}{l} \lambda_{1}(u) =(u+w_1+p)(u+w_2) \\[1.2ex]
    \lambda_{2}(u) =\alpha^2(u+w_1)\end{array}\right. 
 \\[1.2ex]
T^{\Lambda\Lambda}_{jj}(u)\,|0,0> &=& \lambda_{j}(u)\,|0,0>
\ ,\quad j=1,2 \mb{with} \left\{
\begin{array}{l} \lambda_{1}(u) =(u+w_1)(u+w_2) \\[1.2ex]
    \lambda_{2}(u) = \alpha^2\beta^2\end{array}\right. 
\end{eqnarray*}
It implies that the pseudo-vacuum is an eigenvector of the transfer matrix.
We will summarize these different cases in the notation
\begin{equation}
t(u)\,|\Omega> = \Big(\lambda_1(u)+\lambda_2(u)\Big)\,|\Omega>
\end{equation}
where $|\Omega>$ is one of the pseudo-vacuum(s).

The other eigenvalues of the transfer matrix then read 
$$
\lambda(u)=A_0(u)\,\lambda_1(u)+A_1(u)\,\lambda_2(u)
$$
where the dressing functions $A_j(u)$ are given by:
\begin{eqnarray}
A_{0}(u) &=& \prod_{n=1}^{M}
\frac{u-u_n+\half}{u-u_n-\half}\mb{and}
A_{1}(u) = \prod_{n=1}^{M}
\frac{u-u_n-\frac{3}{2}}{u-u_n-\frac{1}{2}}
\end{eqnarray}
The parameters $u_n$, $1\leq n\leq M$, are the Bethe roots, their number
$M$ being also a free parameter. All these parameters are
determined by the Bethe equations 
\begin{eqnarray} 
\prod_{\atopn{m=1}{m\neq n}}^{M} \frac{u_n-u_m-1}{u_n-u_m+1}
=\frac{\lambda_{1}\left(u_n+\frac{1}{2}\right)}
{\lambda_{2}\left(u_n+\frac{1}{2}\right)} \qquad\quad
1 \leq n \leq  M
\end{eqnarray}

\subsection{$N=3$ case}
The existence of a pseudo-vacuum is not ensured anymore. For 
instance in the example treated in section \ref{mixed3}, one easily
computes the action of $t(u)$ on the Fock vacuum:
$$  
t(u)\,|0,0>=\big(u^{2}+\beta_{1}^2\,u +
\beta_{1}^2\,\alpha_{3}^2\big)\,|0,0>
+\beta_{1}^{*}\,\beta_2^{*}\,\alpha_{3}\,b^\dag \,|0,0>
$$
so that the vacuum is an eigenvector of the transfer matrix only if
the condition $\beta_{1}\,\beta_2\,\alpha_{3}=0$ is satisfied. Note
that this condition is weaker than demanding $|0,0>$ to be a lowest
weight vector for the monodromy matrix, which would lead to
$\alpha_{3}=0$.

\subsubsection{Pseudo-vacua \label{sect:pseudovac}}
We give the eigenvalues $\lambda_{j}(u)$ of the pseudo-vacuum (when it exists) under 
the generators $T_{jj}(u)$, $j=1,2,3$. We order them according 
to the section where they are presented.
The pseudo-vacua are built on the Fock vacuum 
$|0>\equiv|0>\otimes\ldots\otimes|0>$ of the models.

\paragraph{Model \ref{sect:BECfield}:} One must impose the condition 
$\beta_{1}\beta_{2}=0$. In that case, there are several pseudo-vacua 
$|p>=(b^\dag)^p\,|0,0,0>$, $p\in\ZZ_{+}$.
\begin{eqnarray}
    T_{11}(u)\,|p> = u^2\,|p> \mb{;}T_{22}(u)\,|p> = \beta_{1}^2(u+p)\,|p> 
    \mb{;}T_{33}(u)\,|p> = \alpha_{3}^2\,\beta_{2}^2\,|p> 
\end{eqnarray}

\paragraph{Model \ref{mixed3}:} One must impose the condition 
$\beta_{1}\beta_{2}=0$. In that case, the Fock vacuum has 
eigenvalues
\begin{eqnarray}
    T_{11}(u)\,|0> = \beta_{1}^2\,u^2\,|0> 
    \mb{;}T_{22}(u)\,|0> = \beta_{2}^2\,u^2\,|0> 
    \mb{;}T_{33}(u)\,|0> = \beta_{3}^2\,u^2\,|0> 
\end{eqnarray}

\paragraph{Model \ref{sect:2cplBEC}:} The Fock vacuum has 
eigenvalues
\begin{eqnarray}
    T_{11}(u)\,|0> = u^2\,|0> 
    \mb{;}T_{22}(u)\,|0> = u^2\,|0> 
    \mb{;}T_{33}(u)\,|0> = \alpha^{2}\,\beta^2\,|0> 
\end{eqnarray}

\paragraph{Model \ref{sect:conform}:} There are two types of pseudo-vacua 
$$
|p> = (a_{2}^\dag)^p\,|0,0,0,0> \mb{and} |-p> = (b_{1}^\dag)^p\,|0,0,0,0>
\ ,\quad p\in\ZZ_{+}
$$
Gathering them in the notation $|p>$, $p\in\ZZ$, their
eigenvalues read
\begin{eqnarray}
    T_{11}(u)\,|p> = \beta^2\,u\,|p> 
    \mb{;}T_{22}(u)\,|p> = u(u+p)\,|p> 
    \mb{;}T_{33}(u)\,|p> = \alpha^{2}\,u\,|p> 
  \ ,\ p\in\ZZ\quad
\end{eqnarray}

\paragraph{Model \ref{sect:glBEC}:} One must impose the condition 
$\alpha\beta=0$. In that case, the Fock vacuum has 
eigenvalues
\begin{eqnarray}
    T_{11}(u)\,|0> = u^2\,|0> 
    \mb{;}T_{22}(u)\,|0> = \beta^2\,u\,|0> 
    \mb{;}T_{33}(u)\,|0> = \alpha^2\,u\,|0> 
\end{eqnarray}

\paragraph{Model \ref{sect:gl3}:} We denote by $v$ the $gl(3)$ lowest 
weight vector, with 
eigenvalues $(\lambda_{1},\lambda_{2},\lambda_{3})$ under the $gl(3)$ 
Cartan generators. The pseudo-vacuum is $|v>=|0>\otimes\,v$, with 
eigenvalues 
\begin{eqnarray}
&& T_{11}(u)\,|v> = u(u+\lambda_{1})\,|v>
 \mb{;}T_{22}(u)\,|v> = u(u+\lambda_{2})\,|v> \\
&& T_{33}(u)\,|v> = \alpha^2\,(u+\lambda_{3})\,|v> 
\end{eqnarray}

\subsubsection{Bethe equations}
Now that we have determined the conditions for the existence of a
pseudo-vacuum, one can write
\begin{equation}
t(u)\,|\Omega> = \Big(\lambda_1(u)+\lambda_2(u)+\lambda_3(u)\Big)\,|\Omega>
\end{equation}
where the eigenvalues $\lambda_{j}(u)$ can be read in the  
section \ref{sect:pseudovac}.
Then, the other transfer matrix eigenvalues read
$$
\lambda(u)=A_0(u)\,\lambda_1(u)+A_1(u)\,\lambda_2(u)+A_2(u)\,\lambda_3(u)
$$
with the dressing functions
\begin{eqnarray}
A_{0}(u) &=& \prod_{n=1}^{M^{(1)}}
\frac{u-u_n^{(1)}+\half}{u-u_n^{(1)}-\half}\\
A_{1}(u) &=& \prod_{n=1}^{M^{(1)}}
\frac{u-u_n^{(1)}-\frac{3}{2}}{u-u_n^{(1)}-\frac{1}{2}}
~~\prod_{n=1}^{M^{(2)}}
\frac{u-u_n^{(2)}}{u-u_n^{(2)}-1}\\
A_{2}(u) &=& \prod_{n=1}^{M^{(2)}}
\frac{u-u_n^{(2)}-{2}}{u-u_n^{(2)}-1}\;.
\end{eqnarray}
We have here two sets of Bethe roots, $u_{n}^{(1)}$, $1\leq n\leq
M^{(1)}$ and $u_{n}^{(2)}$, $1\leq n\leq
M^{(2)}$, constrained by the Bethe equations
\begin{eqnarray}
 \prod_{\atopn{m=1}{m\neq n}}^{M^{(1)}} 
 \frac{u_n^{(1)}-u_m^{(1)}-1}{u_n^{(1)}-u_m^{(1)}+1}\
\prod_{m=1}^{M^{(2)}}
\frac{u_n^{(1)}-u_m^{(2)}+\half}{u_n^{(1)}-u_m^{(2)}-\half}\ =\ 
\frac{\lambda_{1}\left(u_n^{(1)}+\frac{1}{2}\right)}
{\lambda_{2}\left(u_n^{(1)}+\frac{1}{2}\right)} \qquad\quad
1 \leq n \leq  M^{(1)}\\
\prod_{m=1}^{M^{(1)}}
\frac{u_n^{(2)}-u_m^{(1)}+\half}{u_n^{(2)}-u_m^{(1)}-\half}\ 
\prod_{\atopn{m=1}{m\neq n}}^{M^{(2)}} 
\frac{u_n^{(2)}-u_m^{(2)}-1}{u_n^{(2)}-u_m^{(2)}+1}
\ =\ \frac{\lambda_{2}\left(u_n^{(2)}+1\right)}
{\lambda_{3}\left(u_n^{(2)}+1\right)} \qquad\quad
1 \leq n \leq  M^{(2)}
\end{eqnarray}

\subsection{General case}
For the general case of $N\times N$ matrices, and supposing the
existence of pseudo-vacua $|\Omega>$ (possibly with conditions on the 
parameters of the models, as above), we will get for the eigenvalues of
the transfer matrix
$$
\lambda(u) =\sum_{k=1}^{N} A_{k-1}(u)\,\lambda_{k}(u)
$$
with pseudo-vacuum eigenvalues
$$
T_{kk}(u)\,|\Omega> = \lambda_{k}(u)\,|\Omega>\,,\ k=1,\ldots,N
\mb{so that}
t(u)\,|\Omega> = \left(\sum_{k=1}^{N}\lambda_{k}(u) \right)\,|\Omega>
$$
and dressing functions
\begin{eqnarray}
A_k(u)=\prod_{n=1}^{M^{(k)}}
\frac{u-u_n^{(k)}-\frac{k+2}{2}}
{u-u_n^{(k)}-\frac{k}{2}}
~~\prod_{n=1}^{M^{(k+1)}}
\frac{u-u_n^{(k+1)}-\frac{k-1}{2}}
{u-u_n^{(k+1)}-\frac{k+1}{2}}\qquad 0\leq k\leq N-1\;,
\label{dressingClosed}
\end{eqnarray}
with $M^{(0)}=M^{(N)}=0$. 

The $N-1$ types of Bethe roots $u_{n}^{(k)}$, $1\leq n\leq M^{(k)}$,
$1\leq k\leq N-1$ will be determined by the Bethe equations:
\begin{eqnarray}
\label{closedbethe}
&&\prod_{m=1}^{M^{(k-1)}}
\frac{u_n^{(k)}-u_m^{(k-1)}+\half}{u_n^{(k)}-u_m^{(k-1)}-\half}
\ \prod_{\atopn{m=1}{m\neq n}}^{M^{(k)}}
\frac{u_n^{(k)}-u_m^{(k)}-1}{u_n^{(k)}-u_m^{(k)}+1}
\ \prod_{m=1}^{M^{(k+1)}}
\frac{u_n^{(k)}-u_m^{(k+1)}+\half}{u_n^{(k)}-u_m^{(k+1)}-\half}
\ =\ 
\frac{\lambda_{k}\left(u_n^{(k)}+\frac{k}{2}\right)}
{\lambda_{k+1}\left(u_n^{(k)}+\frac{k}{2}\right)} \qquad\quad \nonu
&&\qquad\quad 1\leq n 
\leq M^{(k)} \mb{and} 1\leq k\leq N-1
\end{eqnarray}
They depend on the model and pseudo-vacuum through the eigenvalues 
$\lambda_{k}(u)$, $k=1,\ldots,N$.

\section{Superalgebras and fermions\label{sect:super}}
We have seen that the Yangian is not sufficient to build BEC models based on fermions. 
For such a purpose, one needs to consider another algebraic structure, the 
super-Yangian $Y(M|N)$ based on the superalgebra $gl(M|N)$. 

The models look very similar to the ones presented in section 
\ref{sect:bos}, with the notable difference that the choices for the 
fermionic oscillators is not the same, allowing more flexibility in 
the construction.

\subsection{The superalgebra  $gl(M|N)$}
To define a superalgebra, one needs to introduce a grading $[.]$ 
that will distinguish the
fermionic generators from the bosonic ones. For $gl(M|N)$, 
denoting $e_{ij}$, $i,j=1,...,M+N$ 
these generators, the grading is defined 
as\footnote{Other choices of grading could be 
used, but we will stick to this one throughout the paper.}
\begin{eqnarray}
[e_{ij}]=[i]+[j] \mb{with} [k] = \begin{cases} 0 \mb{if} 1\leq k\leq M \\
1 \mb{if} M+1\leq k\leq M+N \end{cases}
\label{eq:grad}
\end{eqnarray}
The generators with grading 0 are bosonic ones; they form a $gl(M)\oplus gl(N)$
subalgebra of $gl(M|N)$, generated by $e_{ij}$, with $i,j\leq M$ and $i,j>M$.
The remaining generators are of fermionic type, and form a representation 
$(N,\overline{M})\oplus(\overline{N},M)$ of this subalgebra.
The supercommutator reads
$$
[e_{ij}\,,\,e_{kl}\} = -(-1)^{([i]+[j])([k]+[l])}\,[e_{kl}\,,\,e_{ij}\}
=\delta_{jk}\,e_{il} - (-1)^{([i]+[j])([k]+[l])}\,\delta_{il}\,e_{kj}
$$
which amounts to consider commutators when $e_{ij}$ and/or $e_{kl}$ 
are/is bosonic (i.e. of grade 0), and anti-commutators when both are 
fermionic.

The fundamental representation is of dimension $M+N$:
$$ \pi(e_{ij})=E_{ij} $$ where now $E_{ij}$ are graded elementary matrices of size $M+N$.

\subsection{The super-Yangian $Y(M|N)$}
One defines the super-Yangian through a graded $R$-matrix, obeying a graded version of 
YBE. By graded version of YBE, we mean that one has to use a graded tensor product
on the auxiliary spaces:
\begin{equation}
(E_{ij}\otimes E_{kl})\cdot (E_{pq}\otimes E_{rs}) = (-1)^{([k]+[l])([p]+[q])}
\,(E_{ij}E_{pq})\otimes (E_{kl}E_{rs})
\end{equation}
where the grading is the same as the one of $gl(M|N)$.

In fact, everything looks formally the same as for the Yangian, with
the restriction that one has to take care of the grading (\ref{eq:grad}). 
For instance, the $R$-matrix reads:
\begin{equation}
R_{12}(x) =
\II\otimes\II-\frac{1}{x}\,P_{12}, 
\label{Rmat-superY}
\end{equation}
with now the super-permutation operator
$$
P_{12}=\sum_{i,j=1}^{M+N} (-1)^{[j]}\,E_{ij}\otimes E_{ji}
=\sum_{i=1}^{M+N}\left(\sum_{j=1}^{M} E_{ij}\otimes E_{ji}
-\sum_{j=M+1}^{M+N} E_{ij}\otimes E_{ji}\right)\,.
$$
Plugging this $R$-matrix in the relation (\ref{rtt}), and taking care of the graded 
tensor product leads to
\begin{equation} 
 {[L_{ij}(u)\,,\,L_{kl}(v)\}} = \frac{(-1)^{[i]([k]+[l])+[k][l]}}{u-v}
\, \Big(L_{kj}(u)L_{il}(v)-L_{kj}(v)L_{il}(u)\Big)\,.
\end{equation}
We will use elementary monodromy matrices built on $gl(M|N)$:
\begin{equation}
L_{ij}(u) = u\,\delta_{ij} +\, e_{ij}\ ,\ e_{ij}\in\,gl(M|N)
\end{equation}

\subsection{Transfer matrix and symmetries}
The monodromy matrix $\Delta^{(L)} T(u)=T^{[1]}(u)...T^{[L]}(u)$  
gives a transfer matrix of the form
\begin{equation}
st(u)=str(T^{[1]}(u)...T^{[L]}(u)) 
\end{equation}
where the super-trace of a matrix is defined by
\begin{equation}
 str(A)=\sum_{i=1}^{M+N} (-1)^{[i]} A_{ii}
 =\sum_{i=1}^{M} A_{ii}-\sum_{i=M+1}^{M+N} A_{ii}
\mb{for} A=\sum_{i,j=1}^{M+N} A_{ij}\,E_{ij}
\end{equation}
As for the bosonic case, one can show that
\begin{equation}
[st(u)\,,\,st(v)]=0
\end{equation}
Again, to get Hermitian Hamiltonian, we will focus on the case $L=2$, so that
the monodromy and transfer matrices we will be concerned of, have the form
\begin{eqnarray}
T_{kl}(v) &=& \sum_{n=0}^2 v^n\,T_{kl}^{(n)}
\mb{with} T_{kl}^{(2)}=\omega_k\,\delta_{kl}\mb{and}\omega_k\in\CC\\
st(u) &=& t_2\, u^2 + t_1\, u + t_0 \mb{with} t_2\in\CC
\end{eqnarray}
Specializing to $gl(M|N)$ representations will give different models. 
These models will have also a symmetry, as one can see from the relation
\begin{equation}
 {[st(u)\,,\,T_{kl}^{(1)}]} = (\omega_k-\omega_l)\,
(T_{kl}^{(0)}+u\,T_{kl}^{(1)})
\end{equation}
proving again that the quantities
\begin{equation}
I_k=T_{kk}^{(1)}\ ,\ \forall\ k=1,...,N
\end{equation}
commute with the transfer matrix.

We will be essentially interested in the oscillator representation
\begin{equation}
\pi(e_{ij})=a^\dag_i\,a_j 
\mb{with} [a_i\,,\,a^\dag_j\}=\mu_i\delta_{ij}
\end{equation}
where $(a_i,a^\dag_i)$ for $1\leq i\leq M$ (resp. $M+1\leq i\leq M+N$) 
are bosons (resp. fermions), i.e. $[a_i]=[a^\dag_i]=[i]$. 

This choice of grading
implies that $[e_{ij}]=[a^\dag_i]+[a_j]=[i]+[j]$, in accordance with the gradation of
$gl(M|N)$. It corresponds to a bosonic (resp. fermionic)
oscillator representation for the $gl(M)$ (resp. $gl(N)$) subalgebra. Remark that
the opposite choice (i.e. $[a_i]=[a^\dag_i]=[i]+1$) is also possible.

\null

The inhomogeneous oscillator monodromy matrices will then be obtained by taking 
constant \textit{bosonic} oscillators. In that process, the choice of the grading 
for the oscillators will be essential, since it will determine which of the 
oscillators can be possibly set to constant.

\null

Despite of the grading, the elementary matrices are still hermitian.
Hence, the models will be hermitian.

\section{Fermionic BEC models}
\subsection{Two by two (graded) matrices}
We are dealing with the $gl(1|1)$ case, and we focus on the elementary 
monodromy 
matrices of the form
\begin{eqnarray}
\cL(u) &=&
\left(\begin{array}{c|c} u+n_1 & a_1^{\dag}\,a_2 \\ \hline
 a_2^{\dag}\,a_1 & u-n_2 \end{array}\right)
\\
\Lambda(u) &=& 
\left(\begin{array}{c|c} u+n & \beta\,c^{\dag} \\ \hline
\beta\,c & \beta^2 \end{array}\right)
\qquad \wh\Lambda(u) = 
\left(\begin{array}{c|c} -\beta^2 & \beta\,c \\ \hline
\beta\,c^{\dag} &  u-n \end{array}\right)
\end{eqnarray}
where in $\cL(u)$ one of the couple $(a_j,a^\dag_j)$, $j=1,2$,
 is bosonic and the other one fermionic, 
while in $\Lambda(u)$ and $\wh\Lambda(u)$, $(c,c^\dag)$ is fermionic.

\subsubsection{Fermionic heteroatomic BEC model}
We consider $str\cL^{[1]}(u+w_1)\,\cL^{[2]}(u+w_2)$, with 
oscillators $(a_{1},c_{1})$ for $\cL^{[1]}(u)$ and 
$(c_{2},a_{2})$ for $\cL^{[2]}(u)$.
One can consider either $(a_1,a_{2})$ to be bosonic and  $(c_{1},c_2)$
fermionic, or $(a_1,a_{2})$ fermionic  and  $(c_{1},c_2)$ bosonic.

The transfer matrix reads:
\begin{eqnarray}
st(u) &\sim& u(n_{a1}+n_{a2}+n_{c1}+n_{c2})+n_{a1}\,n_{c1}-n_{a2}\,n_{c2}
+w_1(n_{c1}+n_{c2})+w_2(n_{a1}+n_{a2})\nonu
&&+a_1^\dag c_2^\dag c_1  a_{2}+a_2^\dag c_1^\dag c_2 a_1 
\end{eqnarray}
We get a fermionic version of the model described in 
section \ref{sec:simplheteroBEC}.
It can be interpreted as modelizing a coupled pair of one boson and
one fermion which can tunnel together from one well to another.

\subsubsection{Heteroatomic-molecular BEC model with fermions}
We consider $str\cL(u+w_1)\,\Lambda(u+w_2)$, with 
oscillators $(a_{1}\equiv b,a_{2}\equiv a)$ for $\cL(u)$ and 
$c$ for $\Lambda(u)$.
In $\cL(u)$, we take $b$ bosonic. Then, $a$ is fermionic, 
as well as is $c$ in $\Lambda(u)$. We get
\begin{eqnarray}
st(u) &\sim& u(n_{b}+n_{c})+n_{b}\,n_{c}
+w_1\,n_{c}+w_2\,n_{b}+\beta^2\, n_{a}
+\beta\,(a^\dag c^\dag b-b^\dag a c)
\end{eqnarray}
We find a new version of the heteroatomic-molecular BEC model of 
section \ref{sect:atomolBEC}, with now a bosonic molecule $b$ 
constituted of two fermionic atoms $a$ and $c$.

The shift automorphism applied on $b$ produces additional terms in the 
transfer matrix
\begin{eqnarray}
H_{bound} &\sim& \big(\beta\,(b+b^\dag)+\beta^2\big)\,n_{c}
+w_2\,\beta\,(b+b^\dag)
+\alpha\,\beta\,(a^\dag c^\dag - a c)
\end{eqnarray}

\subsubsection{Fermionic two-wells}
We start with $\Lambda^{[1]}(u+w_1)$ and $\Lambda^{[2]}(u+w_2)$, set 
all the bosons to constant, keeping the fermions $c_1$ and $c_2$. 
\begin{eqnarray}
str\Lambda^{[1]}(u+w_1)\,\Lambda^{[2]}(u+w_2) &=& (u+w_1+n_{c1})(u+w_2+n_{c2})
-\beta_1\beta_2\,c_1^\dag c_2+\beta_1\beta_2\, c_1 c_2^\dag  
\end{eqnarray}
\begin{eqnarray}
st(u) &\sim& u(n_{c1}+n_{c2})+n_{c1}\,n_{c2}
+w_1\,n_{c2}+w_2\,n_{c1}
-\beta_1\beta_2\,(c_1^\dag c_2+ c_2^\dag c_1 )
\end{eqnarray}
One recognizes a two-wells models, but now with atoms of fermionic 
nature.

We mention here that there has been recently a great interest in Bose-Einstein condensates with fermions 
since the achievement of quantum degeneracy in ultracold Fermi gases (see, for example 
\cite{reffermi} and references therein).

\subsection{Three by three (graded) matrices}
We are now dealing with the $gl(2|1)$ case, corresponding to $3\times 
3$ matrices. Again the number of possible models becomes numerous, so 
that we present the generic case and treat only two examples, 
physically relevant.

\subsubsection{Generic case}
We present here a general formulation for the transfer matrix, which 
encompasses all the possible models, by setting some of the (bosonic) 
oscillators to constant.

The two generic elementary monodromy matrices have the form
\begin{equation}
L^{[1]}(u) = \left(\begin{array}{cc|c}
u_{1}+n_{1} & a_{1}^\dag\,a_{2} & a_{1}^\dag\,a_{3} \\
a_{2}^\dag\,a_{1} & u_{2}+n_{2} &  a_{2}^\dag\,a_{3} \\
\hline
a_{3}^\dag\,a_{1} &  a_{3}^\dag\,a_{2}  & u_{3}+n_{3}
\end{array}\right)
\mb{and} L^{[2]}(u) = \left(\begin{array}{cc|c}
v_{1}+m_{1} & c_{1}^\dag\,c_{2} & c_{1}^\dag\,c_{3} \\
c_{2}^\dag\,c_{1} & v_{2}+m_{2} &  c_{2}^\dag\,c_{3} \\
\hline
c_{3}^\dag\,c_{1} & c_{3}^\dag\,c_{2}  & v_{3}+m_{3}
\end{array}\right)
\nonumber
\end{equation}
with the notations
\begin{equation}
    n_{j}=a_{j}^\dag\,a_{j} \mb{and} m_{j}=c_{j}^\dag\,c_{j}\qquad 
    j=1,2,3\,.
\end{equation}
For $L^{[1]}(u)$ to be of $gl(2|1)$ type, one can choose either 
$a_{1}$ and $a_{2}$ bosonic and $a_{3}$ fermionic, or 
$a_{1}$ and $a_{2}$ fermionic and $a_{3}$ bosonic. 
Obviously, the same 
criteria apply for $L^{[2]}(u)$, $c_{j}$, $j=1,2,3$.

When the oscillators are bosonic, one can choose to set them to 
constant: in that case the corresponding spectral parameter ($u_{j}$ 
or $v_{j}$) has to be set to zero. In all other cases, it is set to 
$u$. For instance, if $a_{1}$, bosonic, is the only one set to a constant 
$\alpha_{1}$, then, one has $u_{1}=0$ and 
$u_{2}=u_{3}=v_{1}=v_{2}=v_{3}=u$. We will also have in this case 
 $a_{1}^\dag=\alpha_{1}^{*}$ and 
$n_{1}=|\alpha_{1}|^{2}$.

Keeping these rules in mind, one can compute a generic transfer matrix. 
It takes the form (up to irrelevant constant terms)
\begin{eqnarray*}
st(u) &=& \sum_{j=1}^{3} (u_{j}\,m_{j}+v_{j}\,n_{j})\ + H\nonu
H &=&
a_{1}^\dag\,a_{2}\,c_{2}^\dag\,c_{1}+ c_{1}^\dag\,c_{2}\,a_{2}^\dag\,a_{1}+ 
a_{1}^\dag\,a_{3}\,c_{3}^\dag\,c_{1}+ c_{1}^\dag\,c_{3}\,a_{3}^\dag\,a_{1}+ 
a_{2}^\dag\,a_{3}\,c_{3}^\dag\,c_{2}+ c_{2}^\dag\,c_{3}\,a_{3}^\dag\,a_{2}
+\sum_{j=1}^{3} m_{j}\,n_{j}\quad
\end{eqnarray*}
The `true' Hamiltonian $st(0)=H$ of a given model is then obtained 
through the above rules, after 
choosing which of the oscillators are bosonic or fermionic, and, among 
the bosonic ones, which of them are set to constant.

The conserved quantities of the model take the form
$$
u\,I_{j} = u_{j}\,m_{j}+v_{j}\,n_{j}\,,\quad j=1,2,3
$$
Of course, depending of the choices, some of these quantities can be 
zero after use of the rules.

Let us also remark that if one takes all the oscillators to be bosonic 
or fermionic (a situation forbidden in the case of $gl(2|1)$), 
one gets a generic transfer matrix for the usual three by three 
matrices, as 
treated in section \ref{sect:3x3}.

\subsubsection{A fermionic two-coupled BEC}
To exemplify the above techniques, we construct a fermionic version of 
the model given in section \ref{sect:2cplBEC}.

We take $a_{3}$ and $c_{3}$ to be bosonic, and set both of them to 
constant, $\alpha$ and $\gamma$ (both real) respectively.
Then, one gets four couples of oscillators, all of them being fermionic. 
The rules lead to
\begin{eqnarray}
&& u_{1}=u_{2}=v_{1}=v_{2}=u \mb{and} u_{3}=v_{3}=0 \\
&& a_{3}=a_{3}^\dag=\alpha\in\RR \mb{and} n_{3}=\alpha^{2} \\
&& c_{3}=c_{3}^\dag=\gamma\in\RR \mb{and} m_{3}=\gamma^{2} 
\end{eqnarray}
Thus, we get an Hamiltonian
$$
H=a_{1}^\dag\,a_{2}\,c_{2}^\dag\,c_{1}+ c_{1}^\dag\,c_{2}\,a_{2}^\dag\,a_{1}+ 
\alpha\gamma\,(a_{1}^\dag\,c_{1}+ c_{1}^\dag\,a_{1}+ 
a_{2}^\dag\,c_{2}+ c_{2}^\dag\,a_{2})+m_{1}\,n_{1}+m_{2}\,n_{2}
$$
with conserved quantities
$$
I_{1}=n_{1}+m_{1} \mb{and} I_{2}=n_{2}+m_{2}
$$
We recover the Hamiltonian of section \ref{sect:2cplBEC}, with the 
notable difference that the atoms have a fermionic nature.

\subsubsection{Fermionic three coupled BEC model}
If now one takes $a_{3}$ bosonic and constant ($\alpha\in\RR$), and 
$c_{1}$ and $c_{2}$ bosonic and constant ($\gamma_{j}\in\RR$), one 
gets:
\begin{eqnarray}
&& u_{1}=u_{2}=v_{3}=u \mb{and} u_{3}=v_{1}=v_{2}=0 \\
&& a_{3}=a_{3}^\dag=\alpha\in\RR \mb{and} n_{3}=\alpha^{2} \\
&& c_{j}=c_{j}^\dag=\gamma_{j}\in\RR \mb{and} m_{j}=\gamma_{j}^{2} 
    \,,\quad j=1,2
\end{eqnarray}
leading to (using the sign-transposition and dilatation automorphisms 
on $L^{[2]}(u)$)
\begin{eqnarray}
H &=&
\gamma_{1}\gamma_{2}\,(a_{1}^\dag\,a_{2}+ \,a_{2}^\dag\,a_{1})+ 
\alpha\gamma_{1}\,(a_{1}^\dag\,c_{3}+ \,c_{3}^\dag\,a_{1})+ 
\alpha\gamma_{2}\,(a_{2}^\dag\,c_{3}+ c_{3}^\dag\,a_{2})
+\gamma_{1}^{2}\,n_{1}+\gamma_{2}^{2}\,n_{2}-\alpha\,m_{3}
\qquad
\end{eqnarray}
We get the fermionic version of the Hamiltonian described in section 
\ref{mixed3}.

\section{Conclusion}
In the BEC context, we have constructed integrable generalised models in a systematic way
exploring different representations of the $gl(N)$ algebra and the $gl(M|N)$ superalgebra.
Some existing models, such 
as the two-site Bose-Hubbard model, have been recovered and 
 new ones have been predicted.
Interestingly, a two-coupled BEC model with a field, a 
three-coupled BEC model and a two-coupled BEC-model with different
types of atoms, among others, have been introduced. The use of the 
$gl(M|N)$ superalgebra allows the introduction of fermions, leading 
to systems mixing bosons and fermions, as they are presently studied 
in condensed matter BEC experiments. In this context, the `integrable approach' 
can be viewed as a technics to construct in a very general way Hamiltonians relevant 
for these studies.

The energy spectrum of these models has been derived, through the Bethe 
ansatz equations, by the use of analytical Bethe ansatz. The next 
step in the study of these systems in this general framework, is the 
determination of the (Bethe) eigenstates and eigenfunctions, which 
would allow to investigate the classical and quantum dynamics of 
such systems.

Finally, we remark that more general integrable models can be obtained 
using the method presented in the present work. They are constructed 
using products of more elementary monodromy matrices, with the 
restriction that the 
hermiticity of their Hamiltonian is not guaranteed anymore. Apart from 
the trial and error method that one can use on a case-by-case basis, 
a general analysis determining the conditions under which  
Hamiltonians are hermitian would certainly improve the landscape of
integrable BEC models.

\subsection*{Acknowledgments}

A. F. thanks R. R. B. Correia, J. Links and A. P. Tonel for useful discussions.
A. F. also acknowledges support from PRONEX under contract CNPq
66.2002/1998-99 and CNPq (Conselho Nacional de Desenvolvimento
Cient\'{\i}fico e Tecnol\'{o}gico).\\
E.R. wish to thank the Instituto de F\'{\i}sica da Universidade Federal do Rio Grande do Sul
(Porto Alegre) for hospitality during this work was completed.

\appendix
\section{Quantum determinant and conserved quantities}
We have seen that the expansion of the monodromy matrix provides some 
conserved quantities of the integrable models. However,
other conserved quantities can be obtained when considering the center of the 
algebra. For the Yangian, it is known that its center is generated by the 
quantum determinant \cite{MNO}:
\begin{equation}
qdet(u)= \sum_{\sigma\in S_N} sgn(\sigma)\, T_{1,\sigma(1)}(u)\,
T_{2,\sigma(2)}(u-1)\cdots T_{N,\sigma(N)}(u-N+1)
\end{equation}
where $S_N$ is the group of permutations. It is clear that the 
conserved quantities obtained in this way are quite complicated, but 
they may be of some help for the study of the different models.
To illustrate this, we give the form of
the quantum determinant when $N=2$ and 3.
\begin{eqnarray}
qdet(u) &=& T_{11}(u)\,T_{22}(u-1)-T_{12}(u)\,T_{21}(u-1) 
\mb{for} N=2\\
qdet(u) &=& T_{11}(u)\,T_{22}(u-1)\,T_{33}(u-2)
+T_{12}(u)\,T_{23}(u-1)\,T_{31}(u-2)
+T_{13}(u)\,T_{21}(u-1)\,T_{32}(u-2)\nonu
&-& T_{12}(u)\,T_{21}(u-1)\,T_{33}(u-2)
-T_{13}(u)\,T_{22}(u-1)\,T_{32}(u-2)
-T_{11}(u)\,T_{23}(u-1)\,T_{32}(u-2)\nonu
&& \mb{for} N=3
\end{eqnarray}
$qdet(u)$ is a polynomial in $u$, of degree 
$2N$ since $T(u)$ is of degree two. Hence, one gets a priori $2N$ 
conserved quantities. They are not all independent, but they can provide 
new conserved quantities, not contained in the transfer matrix, nor given
by (\ref{eq:consvI}). Remark that these quantities are a priori not hermitian,
but, since they are central in the whole Yangian, so are their adjoint. Hence,
one can build hermitian (and anti-hermitian) conserved quantities from
$qdet(u)$.

Indeed, for $N=2$, expanding $qdet(u)$ from the expansion (\ref{eq:expansT}), 
one gets (up to constant terms)
\begin{eqnarray}
qdet(u) &=& \sum_{n=0}^3 d_n\,u^n\\
d_3 &=& \omega_1\,T_{22}^{(1)}+\omega_2\,T_{11}^{(1)} \\
d_2 &=& \omega_1\,T_{22}^{(1)}-2\omega_2\,T_{11}^{(1)}
+\omega_1\,T_{22}^{(0)}+\omega_2\,T_{11}^{(0)} -T_{12}^{(1)}\,T_{21}^{(1)}\\
d_1 &=& \omega_2\,T_{11}^{(1)}-2\omega_2\,T_{11}^{(0)}
+T_{11}^{(1)}\,T_{22}^{(0)}+T_{11}^{(0)}\,T_{22}^{(1)} 
-T_{12}^{(1)}\,T_{21}^{(0)}-T_{12}^{(0)}\,T_{21}^{(1)}\\
d_0 &=&\omega_2\,T_{11}^{(1)}-T_{11}^{(0)}\,T_{22}^{(1)}
+T_{12}^{(0)}\,T_{21}^{(1)}
\end{eqnarray}
with $\omega_k=\mu_{ak}\,\mu_{bk}$, $k=1,2$, when dealing with 
oscillator representations, or $\omega_{k}=1$ for $gl(N)$ 
representations.

After some algebras, and using the conserved quantities (\ref{eq:consvI}), one
obtains the following invariants
\begin{eqnarray}
c_0 &=& T_{12}^{(0)}\,T_{21}^{(1)}+
T_{11}^{(0)}\,(\omega_2-T_{22}^{(1)})\\
c_1 &=& T_{12}^{(1)}\,T_{21}^{(0)}
+T_{22}^{(0)}\,(\omega_1-T_{11}^{(1)})\\
c_2 &=& T_{12}^{(1)}\,T_{21}^{(1)}-\omega_1\,T_{22}^{(0)}
-\omega_2\,T_{11}^{(0)}
\label{eq:qdet2}
\end{eqnarray}
Of course, the explicit form of these invariant will depend on the 
representations we will use, i.e. on the physical model we are studying.

In the case $N=3$, the same kind of calculation leads to more 
complicated expressions. The simplest ones read:
\begin{eqnarray}
c_4 &=&  \omega_3\,T^{(1)}_{12}\,T^{(1)}_{21} 
+ \omega_2\,T^{(1)}_{13}\,T^{(1)}_{31}
+ \omega_1\,T^{(1)}_{23}\,T^{(1)}_{32}
-\omega_2\,\omega_3\,T^{(0)}_{11} 
- \omega_1\,\omega_3\,T^{(0)}_{22} 
- \omega_1\,\omega_2\,T^{(0)}_{33}
\\
c_0 &=& T^{(0)}_{11}\,T^{(0)}_{22}\,T^{(0)}_{33} 
- T^{(0)}_{11}\,T^{(0)}_{23}\,T^{(0)}_{32} 
- T^{(0)}_{12}\,T^{(0)}_{21}\,T^{(0)}_{33} 
+ T^{(0)}_{12}\,T^{(0)}_{23}\,T^{(0)}_{31} 
+ T^{(0)}_{13}\,T^{(0)}_{21}\,T^{(0)}_{32} 
- T^{(0)}_{13}\,T^{(0)}_{22}\,T^{(0)}_{31}
\nonumber
\end{eqnarray}


\begin{thebibliography}{99}

\bibitem{comp1} M. Batchelor, X-W. Guan, N. Oelkers and A. Foerster,   
\textsl{Thermal and magnetic properties of integrable spin-1 and spin-3/2 chains
 with applications to real compounds}, 
J. Stat. Mech. \textbf{0410} (2004) P017.
  
\bibitem{prl} M. Batchelor, X-W. Guan, N. Oelkers, K. Sakai, Z. Tsuboi and A. Foerster,   
\textsl{Exact results for thermal and magnetic properties of strong coupling ladder compounds}, 
Phys. Rev. Lett.  \textbf{91} (2003) 217202; \\
M. Batchelor, X-W. Guan, N. Oelkers and Z. Tsuboi,
\textsl{Integrable models and quantum spin ladders: comparison between theory 
and experiment for the strong coupling compounds}, 
\texttt{cond-mat/0512489}.
  
\bibitem{rbt} C.T. Black, D.C. Ralph and M. Tinkham, 
Phys. Rev. Lett. \textbf{76} (1996) 688; \\
C.T. Black, D.C. Ralph and M. Tinkham, 
Phys. Rev. Lett. \textbf{78} (1997) 4087; \\
J. von Delft and D. C. Ralph, 
Phys. Rep. \textbf{345} (2001) 61.

\bibitem{baxter1}
R.J. Baxter, 
\textsl{Exactly solved models in statistical mechanics} (Academic Press,
1982).
  
\bibitem{fabkorep} F. H. Essler and V. E. Korepin, 
\textsl{Exactly solvable models of strongly correlated electrons},
(World Scientific, 1994).

\bibitem{bytsko} L. D. Faddeev, 
\textsl{The Bethe Ansatz},
(Andrejeroski Lectures, 1993), SFB288 preprint70.

\bibitem{iachello} F. Iachello and A. Arimo 
\textsl{The Interacting Boson Model}
(Cambridge University Press, 1995).

\bibitem{legget} A. J. Leggett, 
Rev. Mod. Phys. \textbf{73} (2001) 307.

\bibitem{lip}
L.~Lipatov,
\textsl{High Energy Asymptotics of Multi--Colour QCD and Exactly
Solvable Lattice Models,}
JETP Lett. {\bf 59} (1994) 596 and 
\texttt{hep-th/9311037}.

\bibitem{FaKo} L. Faddeev and G. Korchemsky, 
\textsl{High energy QCD as a completely integrable model}
Phys. Lett. \textbf{B342} (1995) 311 and
\texttt{hep-th/9404173}.

\bibitem{Zarembo} G. Ferretti, R. Heise and K. Zarembo,
\textsl{New Integrable Structures in Large-$N$ QCD,}
\texttt{hep-th/0404187}.

\bibitem{BBGK} A.V. Belitsky,  V.M. Braun,  A.S. Gorsky and  G.P.
Korchemsky,
\textsl{Integrability in QCD and beyond},
Int. J. Mod. Phys. \textbf{A19} (2004) 4715 and
\texttt{hep-th/0407232}.

\bibitem{miza}
J.A.~Minahan et K.~Zarembo,
\textsl{The Bethe-Ansatz for N=4 Super Yang-Mills,}
JHEP 0303 (2003) 013 and {\tt hep-th/0212208}.

\bibitem{beisstau}
N.~Beisert et M.~Staudacher,
\textsl{The N=4 SYM integrable super spin chain,}
Nucl. Phys {\bf B670} (2003) 439;\\
\textsl{Long-Range PSU(2,2|4) Bethe Ansaetze for Gauge Theory and
Strings}, Nucl. Phys. \textbf{B727} (2005) 1 and \texttt{hep-th/0504190}

\bibitem{ago} S. N. Bose,
Z. Phys. \textbf{26} (1924) 178; \\
A. Einstein, 
Phys. Math. \textbf{K1 22} (1924) 261.

\bibitem{early} E. A. Cornell and C. E. Wieman, 
Rev. Mod. Phys. \textbf{74} (2002) 875; \\
J. R. Anglin and W. Ketterle, 
Nature \textbf{416}, (2002) 211.

\bibitem{mol} P. Zoller, 
Nature \textbf{417} (2002) 493.

\bibitem{reffermi} J. Hutson and P. Sold\'an, 
\textsl{Molecule formation in ultracold atomic gases}, 
\texttt{physics/0607234}.

\bibitem{hertier} M. H\'eritier,
\textsl{In search of exact solutions}
Nature \textbf{414} (2001) 31; \\
M. T. Batchelor, 
Physics Today \textbf{60} (2007) 36.

\bibitem{transistor} J. Stickney, A. Zozulya and D. Anderson, 
\textsl{Transistor-like Behaviour of 
a Bose-Einstein Condensate in a Triple Well Potential}, \texttt{cond-mat/0607706}.
  
\bibitem{QISM} E.K. Sklyanin and L.D. Faddeev, \textsl{Quantum 
mechanical approach to completely integrable models of field theory},
Dokl. Acad. Nauk. SSSR \textbf{243} (1978) 1430;\\
\textsl{Method of the 
inverse scattering problem and quantum nonlinear Schr\"odinger 
equation}, Dokl. Acad. Nauk. SSSR \textbf{244} (1978) 1337.

\bibitem{QISM2} E.K. Sklyanin, \textsl{Quantum version of the method of 
inverse scattering problem}, Zap. Nauchn. Sem. LOMI \textbf{95} (1980) 
55.

\bibitem{Kul} P.P.~Kulish, N.Yu.~Reshetikhin and E.K.~Sklyanin,
\textsl{Yang-Baxter equation and representation theory: I},
Lett. Math. Phys. \textbf{5} (1981) 393.

\bibitem{KulSkly} P.P.~Kulish and E.K.~Sklyanin,
\textsl{Quantum inverse scattering method and the Heisenberg 
ferromagnet}, Phys. Lett. \textbf{A70} (1979) 461.
    
\bibitem{FRT} L.D.~Faddeev, N.Yu.~Reshetikhin and L.A.~Takhtajan,
\textsl{Quantization of Lie groups and Lie algebras,} Leningrad Math.  J.
\textbf{1} (1990) 193.
 
\bibitem{korepin}
V.E. Korepin, 
\textsl{New effects in the massive Thirring model: repulsive
case}, Comm. Math. Phys. \textbf{76} (1980) 165; \\
V.E. Korepin, G. Izergin and N.M. Bogoliubov, 
\textsl{Quantum inverse
scattering method, correlation functions and algebraic Bethe Ansatz}
(Cambridge University Press, 1993).
  
\bibitem{faddeev}
L.D.~Faddeev and L.A.~Takhtajan, 
\textsl{Spectrum and scattering of
excitations in the one-dimensional isotropic Heisenberg model,}
J. Sov. Math. \textbf{24} (1984) 241; 
\\
L.D.~Faddeev and L.A.~Takhtajan, 
\textsl{What is the spin of a spin wave?} 
Phys. Lett. \textbf{A85} (1981) 375. 

\bibitem{KulResh} P.~Kulish and N.~Yu.~Reshetikhin,   
\textsl{Diagonalisation of $GL(N)$ invariant transfer matrices and quantum 
N-wave system (Lee model)}, J. Phys. \textbf{A16} (1983) L591.

\bibitem{ow} E.~Ogievetsky and P.~Wiegmann,
\textsl{Factorized S-matrix and the Bethe ansatz for simple Lie
groups,} Phys. Lett. \textbf{B168} (1986) 360.

\bibitem{Drinfeld}
V.G.~Drinfel'd, \textsl{Hopf algebras and the quantum Yang--Baxter
equation,} Soviet. Math. Dokl. \textbf{32} (1985) 254; \\
\textsl{A new
realization of {Y}angians and quantized affine algebras,} Soviet. Math.
Dokl. \textbf{36} (1988) 212.

\bibitem{baxter2}
R.J. Baxter, 
\textsl{Partition function of the eight-vertex lattice model,}
Ann. Phys. \textbf{70} (1972) 193; \\ 
J. Stat. Phys. \textbf{8} (1973) 25.

\bibitem{yang}
C.N.~Yang, \textit{Some exact results for the many-body problem in one
dimension with repulsive delta-function interaction},
Rev. Lett. \textbf{19} (1967) 1312.

\bibitem{molinge} Mo-Lin Ge, Kang Xue and Yiwen Wang, 
\textit{An introduction to Yangian in Physics}, Second Pacific 
Winter School in Theoretical Physics, Korea, 1995, \texttt{cond-mat/9509090}.

\bibitem{kundu5} A. Kundu, 
\textsl{Yang-Baxter algebra and generation of quantum integrable models}, 
\texttt{nlin.SI/0609001}.

\bibitem{dukk} G. Ortiz, R. Somma, J. Dukelsky and S. Rombouts,
\textsl{Exactly-solvable models derived from a generalized Gaudin algebra}, 
{Nucl. Phys. \textbf{B707} (2005) 421}.

\bibitem{jc} E. T. Jaynes and F. W. Cummings, Proc. IEEE {\bf 51}, 89 (1963). 

\bibitem{unicamp} D. Jonathan, K. Furuya and A. Vidiella-Barranco, 
\textsl{Dressed-state approach to population trapping in the 
Jaynes-Cummings model}, \texttt{quant-phys/9904067} and references therein.

\bibitem{rego} H. Haken, H. Wolf and W. D. Brewer,
\textsl{The physics of atoms and quanta: Introduction to Experiments and Theory} 
(Springer-Verlag, 2004).

\bibitem{jzrg} J. Links, H.-Q. Zhou, R. H. McKenzie and M. D. Gould, 
J. Phys. \textbf{A 36} (2003) R63; \\
Foerster A, Links J and Zhou H-Q,  
\textsl{Exact solvability in contemporary physics} 
{Classical and Quantum Nonlinear Integrable Systems, 
(I.O.P. Publishing Ltd., 2003}.

\bibitem{maj} M. Duncan, A. Foerster, J. Links, E. Mattei, N. Oelkers and A. Tonel, 
\textsl{Emergent quantum phases in a heteronuclear molecular Bose-Einstein 
condensate model}, 
\texttt{quant-ph/0610244}, to appear in Nuclear Physics B; \\
L. Zhou, W. Zhang, H. Y. Ling and H. Pu, 
\textsl{Quantum correlation in the photoassociation of a heteronuclear Bose-Einstein condensate},
\texttt{quant-ph/0611005}.

\bibitem{milb} G. J. Milburn, J. Corney, E. M. Wright and D. F. Walls,
Phys. Rev.  {\bf A 55} (1997) 4318.

\bibitem{our} A. P. Tonel, J. Links and A. Foerster, J. Phys. 
{\bf A 38} (2005) 1235.

\bibitem{albiez} M. Albiez, R. Gati, J. F\"olling, S. Hunsmann, M. Cristiani
and M.K. Oberthaler, Phys. Rev. Lett. {\bf 95} (2005) 010402.

\bibitem{milb5} K. Nemoto, C. Holmes, G. Milburn and W. Munro, 
Phys. Rev. {\bf A 63} (2000) 013604-1.

\bibitem{chin} B. Liu, L-B. Fu, S-P. Yang and J. Liu, 
\textsl{Josephson Oscillation and Transition to Self-Trapping for Bose-Einstein-Condensates in a 
Triple-Well Trap}, \texttt{cond-mat/0610200}.

\bibitem{4} H. Ott et al, Phys. Rev. Lett. {\bf 87} (2001) 230401; W. H\"ansel et al, Nature {\bf 413} (2001) 
498; J. Reichel, Apl. Phys. {\bf B 75} (2002) 469.

\bibitem{prlpena} P. Buonsante, R. Franzosi and V. Pena, Phys. Rev. Lett. {\bf 90} (2003) 050404

\bibitem{tilted} D.R. Dounas-Frazer, A. M. Hermundstad and L. D. Carr, 
\textsl{Ultracold bosons and entanglement 
in the tilted double-well}, \texttt{quant-ph/0609119}.

\bibitem{byebye} D. Arnaudon, N. Cramp\'e, A. Doikou, L. Frappat, E. Ragoucy,
\textsl{Analytical Bethe Ansatz for closed and open $gl(n)$-spin chains in
any representation,} JSTAT \textbf{02} (2005) P02007,
\texttt{math-ph/0411021}.

\bibitem{MNO}
A.~Molev, M.~Nazarov and G.~Olshanski, 
\textsl{Yangians and classical Lie
algebras}, Russian Math. Survey \textbf{51} (1996) 205 and
\texttt{hep-th/9409025}.

\end{thebibliography}
\end{document}